\documentclass[fleqn,usenatbib]{mnras}

\usepackage[T1]{fontenc}
\usepackage{ae,aecompl}

\usepackage{graphicx}	
\usepackage{amsmath}	
\usepackage{amssymb}	

\title[3D and flickering mapping of V4140\,Sgr]
 {SOAR observations of the high-viscosity accretion disc \\ of the
  dwarf nova V4140~Sagitarii in quiescence and in outburst
\thanks{Based on observations obtained at the Southern Astrophysical
  Research (SOAR) telescope, which is a joint project of the 
  Minist\'{e}rio da Ci\^{e}ncia, Tecnologia, e Inova\c{c}\~{a}o
  (MCTI) da Rep\'{u}blica Federativa do Brasil, the U.S. National
  Optical Astronomy Observatory (NOAO), the University of North
  Carolina at Chapel Hill (UNC), and Michigan State University
  (MSU).}}

\author[Baptista et~al.]
 {Raymundo Baptista$^{1}$\thanks{E-mail: raybap@gmail.com},
  Bernardo W. Borges$^{2}$ \& Alexandre S. Oliveira$^{3}$
\\
$^{1}$Departamento de  F\'{\i}sica, Universidade  Federal de
Santa Catarina, Trindade, 88040-900, Florian\'{o}polis, SC, Brazil\\ 
$^{2}$Universidade Federal de Santa Catarina, Campus Ararangu\'a,
CEP 88905-120, Ararangu\'a, SC, Brazil\\
$^{3}$IP\&D, Universidade do Vale do Para\'{\i}ba, CEP 12244-000,
S\~ao Jos\'e dos Campos, SP, Brazil\\
}

 \date{Accepted 2016 September 5. Received 2016 September 5;
   in original form 2016 January 26}

\pubyear{2016}

\begin{document}
\label{firstpage}
\pagerange{\pageref{firstpage}--\pageref{lastpage}}
\maketitle

\begin{abstract} 

We report the analysis of 22 B-band light curves of the dwarf nova
V4140~Sgr obtained with SOI/SOAR during two nights along the decline
of a superoutburst in 2006 Sep 12-24 and in quiescence over 50 days
following the superoutburst. Three-dimensional eclipse mapping
of the outburst light curves indicates that the accretion disc is
elliptical (eccentricity $e=0.13$) and that superhump maximum
occurs when the mass donor star is aligned with the bulge of the
elliptical disc. The accretion disc is geometrically thin both in
outburst and in quiescence; it fills the primary Roche lobe in
outburst and shrinks to about half this size in quiescence.
The stability of the eclipse shape, width and depth along
quiescence and the derived disc surface brightness distribution
indicate that the quiescent accretion disc is in a high-viscosity,
steady-state.
Flickering mapping of the quiescent data reveal that the low-frequency
flickering arises from an azimuthally-extended stream-disc impact
region at disc rim and from the innermost disc region, whereas the
high-frequency flickering originates in the accretion disc.
Assuming the disc-related flickering to be caused by fluctuations
in the energy dissipation rate induced by magneto-hydrodynamic
turbulence \citep{ga}, we find that the quiescent disc viscosity
parameter is large $\alpha\simeq 0.2-0.4$ at all radii.
The high-viscosity quiescent disc and the inferred low disc
temperatures in superoutburst are inconsistent with expectations
of the disc-instability model, and lead to the conclusion that
the outbursts of V4140~Sgr are powered by mass transfer bursts
from its donor star.

\end{abstract}


\section{Introduction}

In dwarf novae, mass is transferred from a late-type star (the
secondary) to a companion white dwarf (WD) via an accretion disc. 
They show recurrent outbursts on timescales of days-months, in 
which the disc brightens by factors 20-100 for $\simeq 1-10$ days. 
Outbursts are explained in terms of either a thermal-viscous 
disc-instability model \cite[DIM, e.g.,][]{c93,lasota} or a 
mass-transfer instability model \cite[MTIM, e.g.,][]{bath}. 
DIM predicts matter accumulates in a low viscosity
\footnote{here we adopt the prescription of \cite{ss} for
  the accretion disc viscosity, $\nu = \alpha c_s H$, where
  $\alpha$ is the non-dimensional viscosity parameter, $c_s$ is
  the local sound speed and $H$ is the disc scaleheight.}
disc during quiescence ($\alpha_{\rm quies}\sim 10^{-2}$) which
switches to a high-viscosity regime during outbursts, whereas 
in MTIM the disc viscosity is always the same
\cite[$\alpha\sim 0.1-1$, from the decline timescale of
outbursting dwarf novae, e.g.,][]{w95}. Therefore, estimating
$\alpha$ of a quiescent disc is key to gauge which model is at
work in a given dwarf nova.

Aside of the normal outbursts, the short-period dwarf novae of the
SU\,UMa type show longer, slightly brighter and more regular
superoutbursts, during which a hump-shaped brightness modulation
(named {\em superhump}) with period slightly longer than the orbital
period $P_\mathrm{orb}$ is seen in their light curves \citep{w95,hellier}.
The most promising explanation for superhumps is given by the tidal
resonance instability model \citep{whitehurst88,ho90,lubow94}: during
a superoutburst, the accretion disc expands beyond the 3:1 resonance
radius, $R_{31}$, and a tidal instability sets in, giving rise to an
elliptical, slowly precessing disc.
Superhump modulation then arises from the periodic tidal interaction
between the outer elliptical disc and the mass-donor star (normal
superhumps) or by the varying release of gravitational energy at the
point where the gas stream hits the precessing elliptical disc outer
edge (late superhumps), at the beat period between the orbital and
the disc precession period, $1/P_\mathrm{sh}= 1/P_\mathrm{orb} -
1/P_\mathrm{prec}$.  Eclipse mapping \citep{horne85} is a powerful
tool to test for the presence of elliptical discs in outbursting
dwarf novae as well as to check whether the orientation of the
ellipse is according to the expectation of the tidal instability
model \cite[e.g.,][]{odonoghue,rolfe00}.

Flickering is the intrinsic brightness fluctuation of 0.01-1 mag on
timescales of seconds to dozens of minutes seen in dwarf novae light
curves. Optical studies \citep[e.g.,][]{bruch92,bruch96,bruch00,bb04}
 suggest there may be three different sources of flickering in
 dwarf novae and novalike systems, the relative importance of which
 varies from system to system: (i) the stream-disc impact region
 \cite[possibly because of unsteady mass transfer or post-shock
 turbulence,][]{wn,shu76}, (ii) the innermost disc regions around
 the WD \cite[possibly powered by unsteady WD accretion or
 post-shock turbulence in the boundary layer between disc and WD,]
 []{ej82,bruch92}, and (iii) the accretion disc itself
 \cite[probably as a consequence of magneto-hydrodynamic (MHD)
 turbulence or events of magnetic reconnection at the disc
 atmosphere,][]{ga,kawa,bb04}. The power density spectrum (PDS)
 of the flickering is characterized by a continuum power-law, $P(f)
 \propto f^{-n}$, with $n\simeq 1-3$ \citep{bruch92}, which flattens
 below a given cut-off frequency $f_c$.

 Currently, the most promising explanation for the anomalously large
 viscosity of accretion discs is related to MHD turbulence in the
 differentially rotating disc gas \citep[driven by the magnetorotational
 instability, MRI, see][]{bh91,hb91}. Most of the studies on this
 subject over the last two decades focused on confirming that MRI
 leads to self-sustained, turbulent and efficient outward flow of
 angular momentum and inward flow of disc matter \citep[e.g.,][and
 references therein]{bh98,beckwith11}, and on the comparison of
 the numerically derived values of $\alpha$ with those inferred by
 the decline timescale of outbursting dwarf novae
 \citep[e.g.,][]{king07,hirose14}.
 On the other hand, the study of \cite{ga} focused on the time
 variability of the viscous energy release in a MHD turbulent disc.
 They found that MHD turbulence leads to large fluctuations in the
 energy dissipation rate per unit area at the disc surface, $D(R)$,
 which they suggested could be a source of flickering in
 mass-exchanging binaries. Indeed, the PDS of the fluctuations in
 their study resemble those of flickering sources, with a power-law
 dependency of similar index range and a flat slope at low-frequencies.
 The interpretation of the fluctuations in $D(R)$ as the stochastic
 and statistically independent release of energy from a large number
 of turbulent eddies leads to a direct relation between the relative
 amplitude of the energy fluctuations $\sigma_\mathrm{D}/\langle D \rangle$
 and the disc viscosity parameter, $\sigma_\mathrm{D}/\langle D \rangle
 \propto \alpha^{1/2}\,(H/R)^{1/2}$, providing an interesting
 observational way to estimate the local accretion disc viscosity
 parameter \cite[see, e.g.,][]{bb04}. For a thin accretion disc
 ($H\simeq 0.01\,R$), the above relation predicts that low-viscosity
 discs ($\alpha \sim 10^{-2}$) should show low-amplitude disc
 flickering (hardly detectable at a level $\leq 1$ per cent),
 whereas high-viscosity discs ($\alpha=$ 0.1-1) should display
 detectable flickering with relative amplitudes in the range
 2.5-7.5 per cent. There is observational support for this
 prediction: e.g., the accretion disc seems the dominant source of
 flickering in the high-viscosity discs of the novalike systems
 RW\,Tri \citep{hs85} and UX\,UMa \citep{bruch00}, whereas there
 is no evidence of disc-related flickering in the low-viscosity
 accretion discs of the dwarf novae U\,Gem \citep{wn} and IP\,Peg
 \citep{bruch00} in quiescence.

V4140 Sgr is an 88-min period eclipsing SU~UMa type dwarf nova
\citep{js,bjs89} showing low-amplitude ($\sim 1$ mag), 5-10\,d long
outbursts recurring every 80-90\,d and longer, brighter superoutbursts
where superhumps appear in its light curve \citep{bb05}.
Here we report the analysis of a sample of light curves of V4140~Sgr
with eclipse mapping techniques to trace the evolution of the surface
brightness of its accretion disc during decline from a superoutbursts,
to locate the sources of flickering in the binary and to estimate
the radial run of the quiescent disc viscosity parameter.
Section\,\ref{observa} reports the observations and data reduction
procedures. Data analysis and results are presented in 
section\,\ref{analysis}, discussed in section\,\ref{discuss},
and summarized in section\,\ref{conclusions}.

\section{Observations and Data Reduction} \label{observa}

Time series of B-band CCD photometry of V4140~Sgr were obtained
along 2006 with the SOAR Optical Imager (SOI) at the 4.1\,m SOAR
telescope, in Cerro Pach\'on, Chile. The SOI camera has a 
mini-mosaic of two back illuminated E2V $2048 \times 4096$ pixels 
CCDs covering a 5.26 arcminute square field of view. The SOI 
detectors are mounted with their long sides parallel and spaced 
102 pixels apart, resulting in a $8\arcsec$ gap between the 
individual CCD images. All observations were performed in its 
$4 \times 4$ pixels binning and fast-readout mode (6.1\,s total
readout time) at a spatial scale of $0.31\arcsec\,{\rm pixel}^{-1}$.
The observations are summarized in Table~\ref{tab1}.
Column 3 lists the number of points in the light curve ($N_p$),
column 4 gives the exposure time in seconds ($\Delta t$). 
Column 5 lists the eclipse cycle number (E); observations that do 
not cover the eclipse are indicated in parenthesis, and those with
incomplete eclipse phase coverage are indicated by a colon after
the cycle. Column 7 gives an estimate of the quality of each run. 
The observations comprise 22 light curves obtained with the same
instrument and telescope, ensuring a high degree of uniformity to
the data set. 
%
\begin{table*}
 \begin{minipage}{120mm}
   \centering
   \caption{Journal of the observations} \label{tab1}
   \begin{tabular}{ccrccccc}
\hline \noalign{\smallskip}
Date (UT) & HJD start & $N_p$ & $\Delta\,t$ &
E \footnote{With respect to the ephemeris of eq.~(\ref{efem}).} &
Phase & Night & Seeing \\ [-0.5ex]
(2006) & (2.450.000+) && (s) & (cycle) & Range &
Quality \footnote{Sky conditions: (A) photometric, (B) good,
(C) poor (large variations, clouds, or both).} & ($\arcsec$) \\
\noalign{\smallskip}
\hline \noalign{\smallskip}
Sep 12 & 3991.53678 & 528 & 3.7 & 125833 & $-0.50,+0.50$ & A & 0.8-2.0 \\
  "    & 3991.58268 & 446 & 3.7 & 125834 & $-0.50,+0.50$ & B \\
  "    & 3991.64405 & 467 & 3.7 & 125835 & $-0.50,+0.34$ & B \\
Sep 24 & 4003.60079 & 205 & 3.7 &(126029)& $+0.14,+0.50$ & A & 0.7-1.0 \\
  "    & 4003.62291 & 548 & 3.7 & 126030 & $-0.50,+0.50$ & B \\
  "    & 4003.68432 & 385 & 3.7 & 126031 & $-0.50,+0.20$ & B \\
Sep 27 & 4006.50008 &  77 & 3.7 &(126076)& $+0.34,+0.50$ & C & 0.7-0.9 \\
  "    & 4006.51006 & 547 & 3.7 & 126077 & $-0.50,+0.50$ & B \\
  "    & 4006.57151 & 557 & 3.7 & 126078 & $-0.50,+0.50$ & B \\
  "    & 4006.64293 & 548 & 3.7 & 126079 & $-0.50,+0.50$ & B \\
  "    & 4006.69433 & 270 & 3.7 & 126080:& $-0.50,-0.02$ & B \\
Sep 29 & 4008.49251 & 428 & 3.7 & 126109 & $-0.23,+0.76$ & A & 0.7-0.9 \\
Oct 01 & 4010.54703 &  95 & 5.0 &(126142)& $+0.29,+0.50$ & B & 0.8-1.1 \\
  "    & 4010.56445 & 442 & 5.0 & 126143 & $-0.50,+0.50$ & B \\
  "    & 4010.62589 & 207 & 5.0 & 126144:& $-0.50,-0.03$ & B \\
Oct 21 & 4030.48861 & 359 & 4.0 & 126467 & $-0.16,+0.52$ & A & 0.7-0.8 \\
Nov 14 & 4054.52175 & 202 & 4.0 &(126858)& $+0.07,+0.50$ & A & 0.8-1.0 \\
  "    & 4054.54806 & 499 & 4.0 & 126859 & $-0.50,+0.44$ & A \\
Nov 15 & 4055.51412 & 131 & 4.0 &(126874)& $+0.23,+0.50$ & A & 0.8-1.1 \\
  "    & 4055.53092 & 405 & 4.0 & 126875 & $-0.50,+0.26$ & A \\
Nov 16 & 4056.51947 & 370 & 4.0 & 126891 & $-0.41,+0.29$ & A & 0.9-1.2 \\
Nov 17 & 4057.51611 & 359 & 4.0 & 126907 & $-0.18,+0.51$ & B & 1.0-1.6 \\
\noalign{\smallskip}
\hline \\ [-8ex]
    \end{tabular}
 \end{minipage}
\end{table*}

Data reduction procedures included bias subtraction, flat-field 
correction, cosmic rays removal and aperture photometry extraction. 
Time series were constructed by computing the magnitude difference 
between the variable and a bright reference comparison star 
$60\arcsec$~S of the variable \citep[star C1 in the finding chart
of][]{bb05} with scripts based on the aperture photometry 
routines of the apphot/IRAF package
 \footnote{IRAF is distributed by National Optical Astronomy 
 Observatories, which is operated by the Association of 
 Universities for Research in Astronomy, Inc., under contract with
 the National Science Foundation.}.  
Light curves of other comparison stars in the field were also 
computed in order to check the quality of the night and the 
internal consistency and stability of the photometry over the 
time span of the observations.  
The magnitude and colors of the reference star were tied to the 
Johnsons-Cousins BVRI system \citep{bessell} from observations of
this star and of standard stars \citep{graham,sb} made on a 
photometric night. The reference star has a calibrated magnitude
of $\mathrm{B}= 14.75 \pm 0.05$~mag. We used the relations of 
\cite{lamla} to transform its $B$-band magnitude to flux units and
to convert the light curves of the variable star from magnitude
difference to absolute flux with an estimated photometric accuracy
of 5 per cent. Moreover, the analysis of the light curves of 
field comparison stars of brightness comparable to that of the
variable indicates that the internal error of the photometry is 
less than 2 per cent.  The error in the photometry of the variable
is derived from the photon count noise and is transformed to flux
units using the same relation applied to the data. The individual
light curves have typical signal-to-noise ratios of $S/N=50$
out-of-eclipse and $S/N=10$-20 at mid-eclipse.

\begin{figure}
\includegraphics[scale=0.34,angle=270]{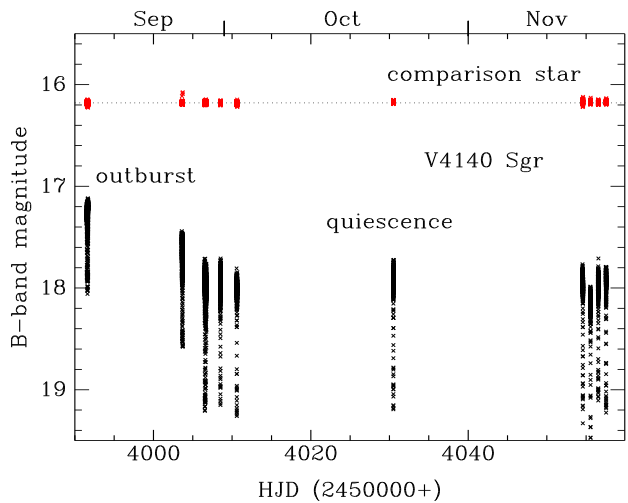}
\caption{Combined SOAR light curves of V4140~Sgr (black crosses)
 for the epoch 2006 September to November. Excursions to higher
 magnitudes indicate eclipses. Measurements of a brighter 
 comparison star are shown in red for illustrations purposes. }
\label{fig1}
\end{figure}

Fig.\,\ref{fig1} shows the 2006 observations of V4140~Sgr.
%
%
It was caught in outburst in Sep 12 ($\simeq 1$\,mag brighter
than in quiescence), while the Sep 24 observations frame the 
decline from that outburst. The object was back to its
quiescent state from Sep 27 onwards. Given the typical shape
and length of the V4140~Sgr outbursts \citep{bb05}, the Sep 12
data probably correspond to the outburst plateau phase
(i.e., past outburst maximum).
As previously noted by \cite{bb05}, the eclipses in outburst are 
shallower than in quiescence, indicating that the brightness 
increase is mainly from the outer and partially eclipsed disc 
regions. The presence of superhumps in the outburst light curves 
(Sect.\,\ref{super}) indicates that this was a superoutburst.

\section{Data analysis and Results} \label{analysis}

\subsection{The Light Curves} \label{lcurve}

The light curves were phase-folded according to the linear plus
sinusoidal ephemeris of Baptista et~al (2003),
\begin{eqnarray}
T_{mid} = \mathrm{BJDD}\,\, 2\,446\,261.671\,50 \nonumber \\
\mbox{} + 0.061\,429\,6757 \times E \nonumber \\
 \mbox{} + 19\times 10^{-5}\,\cos\left[2\pi\left( 
\frac{E- 3\times 10^3}{41\times 10^3}\right) \right] \, d \, .
\label{efem}
\end{eqnarray}
where $E$ is the cycle number and $T_{mid}$ gives the WD 
mid-eclipse times. We checked whether this ephemeris correctly 
centred the WD eclipse by median combining all quiescent data, 
computing the derivative of the combined light curve, and measuring
the phases of minimum/maximum in the derivative on the assumption
that they indicate the mid-ingress/egress phases of the WD eclipse.
This leads to a WD eclipse width of $\Delta\phi = 0.0384 \pm 
0.0006$ cycle, in good agreement with the value derived by
\cite{bjs89}. We also find a WD mid-eclipse phase of $\phi_0=
+0.0039$ cycle, indicating that the observed eclipses occurred
 20.5\,s after the prediction of the ephemeris of 
Eq.(\ref{efem}). All light curves were then offset by $-0.0039$
cycle to make the WD eclipse centre coincident with phase zero.

\begin{figure}
\includegraphics[angle=270,scale=0.34]{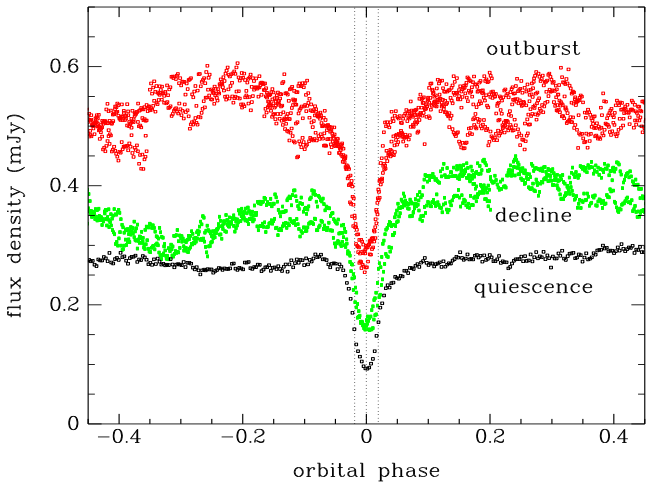}
\caption{Phase-folded light curves of V4140~Sgr for Sep 12 ({\it 
 outburst}, red open squares) and Sep 24 ({\it decline}, green
 filled squares), together with the median of the 16 individual 
 light curves in quiescence ({\it quiescence}, black open squares). 
 Vertical dotted lines mark mid-eclipse and WD ingress/egress 
 phases. }
\label{fig2}
\end{figure}

Fig.\,\ref{fig2} shows phase-folded light curves of V4140 Sgr
grouped per brightness state.
%
%
The median quiescent light curve displays a steep, narrow eclipse
with a long egress shoulder and a weak orbital hump with an apparent
maximum at phase $-0.08$ cycle, signaling a small contribution from
the bright spot at disc rim to the total flux.
The superoutburst light curves show strong flickering (at $\simeq 
20$ per cent level) and a pronounced, broad superhump centred at 
phase zero. In the decline light curves, flickering amplitude is 
reduced (in absolute and relative terms) and the superhump maximum
is displaced to phase $\simeq +0.25$ cycle. Flares and 
flickers occurring right before ingress and just after egress of
the WD suggest that a significant fraction of the flickering in 
superoutburst and decline arises from the inner disc regions. 
Unfortunately, the small number of light curves covering 
superoutburst and decline prevents us from applying flickering 
mapping techniques (see Sect.\,\ref{quies}) to derive the spatial
distribution of the flickering sources at these brightness states.
The total width of the eclipses decreases with decreasing brightness 
level suggesting that the accretion disc shrinks along the decline
from superoutburst, being smallest in quiescence.
Furthermore, the depth of the eclipses increases with decreasing
brightness signaling that the relative contribution of 
the outer disc regions to the total flux decreases along the 
decline and is also smallest in quiescence.

\subsection{On the Nature of the Outburst} \label{super}

Was the 2006 Sep outburst a normal one or a superoutburst?
A straightforward way to address this questions is by searching
for superhumps in the outburst and decline light curves.
In order to detect non-orbital periodicities, we removed the 
eclipses (phase range $-0.1$ to $+0.1$) from the light curves and 
computed Lomb-Scargle periodgrams \citep{numrec} separately for
the Sep 12 and Sep 24 data. 
The power spectrum of the latter data set shows a periodicity at 
frequency $f_\mathrm{sh}= 14.9 \pm 1.0\, \mathrm{cycle\; d}^{-1}$,
corresponding to a period of $P_\mathrm{sh}= 0.067\pm 0.005$~d, 
where the quoted uncertainty is derived from the width at half 
power of the observed peak. The same periodicity is found in the
Sep 12 periodgram, although the corresponding power spectrum 
peak is less pronounced in this case. 

This period is consistently longer than the orbital period,
and is the same in both nights within the uncertainties.
 Moreover, the maximum of the observed modulation coincides with
 phase zero on Sep 12 and has shifted to phase $\simeq +0.25$ cycle
 on Sep 24 (Fig.\ref{fig2}), indicating that it is not the consequence
 of anisotropic emission from a bright spot at disc rim, and confirming
 that $P_\mathrm{sh}$ is different from the orbital period. Also,
 the length of the 2006 Sep outburst is at least 13 days (it probably
 started before Sep 12), which is clearly longer than the typical
 5-10\,d duration of the normal outbursts of this dwarf nova
 \citep{bb05}.
These evidences give us confidence that the observed modulations in
Sep 12 and Sep 24 light curves are indeed superhumps and, therefore,
that these observations framed a superoutburst of V4140~Sgr.

As a consequence of the short time span of the observations (only 
2-3 binary orbits in each date), the uncertainty in $P_\mathrm{sh}$
is too large to allow a meaningful derivation of a superhump 
period excess from $P_\mathrm{sh}$. We attempted to increase
the accuracy of the measurement of $P_\mathrm{sh}$ by computing
a power spectrum of the combined light curve of both nights.
However, no periodicity other than harmonics of the orbital
period appears in this case, suggesting a significant change in
phase of the superhump signal along this time interval -- as is
usual in superoutbursting dwarf novae \citep[e.g.,][]{patterson00}.

\subsection{3D Eclipse Mapping in Outburst} \label{erup}

We grouped the data per outburst stage to generate phase-binned,
median-averaged light curves for the 'outburst' (Sep 12 data), 
'decline' (Sep 24 data) and 'quiescence' (the remaining data sets)
brightness states. The resulting light curves have phase resolutions
of 0.003~cycle during eclipse (phase range $-0.1$ to $+0.1$ cycle)
and of 0.009~cycle outside of eclipse. Only orbital phases in the
range $-0.4$ to $+0.4$~cycle are included because our eclipse
mapping algorithm does not handle a possible eclipse of the 
secondary star by the accretion disc (at phase $\pm 0.5$).

Three-dimensional (3D) maximum entropy eclipse mapping techniques 
were applied to these light curves in order to investigate the
source of the superhump and to follow the evolution of the disc
surface brightness distribution along the superoutburst decline. 
The reader is referred to \cite{horne85} for the historical 
presentation, \cite{rutten98} for a description and performance
tests of the 3D eclipse mapping algorithm, and \cite{bap01} for a
more recent review on eclipse mapping techniques.

Our 3D eclipse map consists of a grid of $51\times 51$ pixels on a
conical surface with side $2\,R_\mathrm{L1}$ (where $R_\mathrm{L1}$
is the distance from disc centre to the inner Lagrangian point)
centred at the WD position and inclined at a half-opening angle
$\beta$ with respect to the orbital plane, plus a circular rim of 
102 pixels orthogonal to the orbital plane at a distance $R_d$ 
($< R_\mathrm{L1}$) from the disc centre.
The disc rim allows us to model out-of-eclipse modulations (such as
anisotropic emission from the bright spot at stream-disc impact) 
as the fore-shortening of an azimuthally-dependent brightness
distribution in the disc rim pixels \citep{bob97,tiago07}.
The eclipse geometry is defined by the mass ratio $q$ and the 
inclination $i$, and the scale of the map is set by $R_\mathrm{L1}$.
We adopted $R_\mathrm{L1}=(0.427\pm 0.035)R_\odot$, $q=0.125\pm 0.015$
and $i=80.2\degr \pm 0.5\degr$ \citep{bb05}, which correspond to a WD
eclipse width of $\Delta\phi = 0.0378\pm 0.0005$~cycle \citep{bjs89}.
This combination of parameters ensures that the WD is at the centre
of the map.  

The light curves were analyzed by the 3D eclipse mapping algorithm
to solve for a map of the disc + rim surface brightness distribution.
The reconstructions were performed with a polar Gaussian default
function \citep{rutten92} with radial blur width $\Delta r= 0.02\;
R_{L1}$ and azimuthal blur width $\Delta\theta= 30\degr$, and 
reached a final reduced chi-square $\chi_\mathrm{red}^{2}\simeq 1$
for all light curves.  The uncertainties in the eclipse maps were 
derived from Monte Carlo simulations with the light curves using 
the bootstrap technique \citep{boots,wd01}, generating a set of 20 
randomized eclipse maps \citep[see][]{rutten92}. These are combined
to produce a map of the standard deviations with respect to the 
true map. A map of the statistical significance (or the inverse of
the relative error) is obtained by dividing the true eclipse map by 
the map of the standard deviations \citep{b05}.
The uncertainties obtained with this procedure are used to draw the 
contour maps of Fig.\,\ref{fig4}, and to estimate the uncertainties 
in the derived disc rim intensity (Fig.\,\ref{fig4}) and 
radial brightness temperature distributions (Fig.\,\ref{fig5}).

\subsubsection{Disc rim and disc half-opening angle} \label{discrim}

An entropy landscape technique was used to derive the best-fit disc 
rim radius $R_d$ and half-opening angle $\beta$ for each brightness 
state (see Appendix). We obtained reconstructions for disc rims 
in the range $(0.35-0.85)\,R_\mathrm{L1}$ at steps of $0.05\,R_
\mathrm{L1}$, and half-opening angles in the range $(0\degr - 6\degr)$ 
at steps of $0.5\degr$,
  and investigated the resulting space of parameters in search
  of the combination $(R_d,\beta)$ that yields the eclipse map of
  highest entropy. The space of parameters is well behaved with a
  single, well defined entropy maximum in each case.
$R_d$ is best constrained in outburst, while $\beta$ is better 
constrained in quiescence: the entropy of the eclipse maps drops 
quickly for $R_d<0.75\, R_\mathrm{L1}$ in outburst, and for 
$\beta>1.0^o$ in quiescence.
The accretion disc fills the primary Roche lobe at outburst 
($R_d=0.85\,R_\mathrm{L1}\,=0.36\,R_\odot$). It shrinks to $0.6\,
R_\mathrm{L1}\,(=0.25\,R_\odot)$ during decline and reaches 
$0.45\,R_\mathrm{L1}\,(=0.19\,R_\odot)$ in quiescence
\footnote{ The uncertainty is $0.05\,R_\mathrm{L1}=0.021\,R_\odot$
  in each case.}.
These results are in line with the qualitative inferences drawn
in Sect.\,\ref{lcurve} and in agreement with the $\simeq 0.4
\,R_\mathrm{L1}$ quiescent disc radius found by \cite{bb05}. 
There is marginal evidence that the disc half-opening angle is
larger in outburst ($\beta= 1.0^o\pm 0.5^o$) than in quiescence
($\beta= 0.5^o\pm 0.5^o$). The disc is geometrically thin in all
three cases.

In order to check the consistency of the inferred $R_d$ values, 
we applied the method described by \cite{sulkanen} to estimate
the disc radius from the half-width of the eclipse $\Delta\phi_E$.
By assuming a spherical secondary star it is possible to
derive the disc radius $R_d$ in units of the orbital separation 
$a$ from the analytical expression,
\begin{equation}
\frac{R_d}{a} = 
\sin(2\pi\Delta\phi_E)\sin i - \sqrt{(R_2/a)^2 - \cos^2 i} \,,
\end{equation}
where $R_2/a$ is the radius of a sphere containing the same 
volume as the Roche lobe of the secondary star, given by the 
relation \citep{eggleton},
\begin{equation}
\frac{R_2}{a}= \frac{0.49\,q^{2/3}}{0.6\,q^{2/3}+\log(1+q^{1/3})}\, .
\end{equation}
Values of $R_d/a$ are transformed to $R_d/R_\mathrm{L1}$ assuming
$a/R_\odot= 0.61\pm 0.05$ \citep{bb05}.
Measurements of $\Delta\phi_E$ were obtained by fitting a spline
function to the out-of-eclipse phases of each light curve,
finding the phases where the fitted spline deviates from the data
by 1$\sigma$ of the local, out-of-eclipse median flux, and 
dividing the resulting phase range by two. We find $\Delta\phi_E$
values of $0.122\pm 0.005$~cycle, $0.097\pm 0.005$~cycle and 
$0.076\pm0.005$~cycle, respectively for the outburst, decline and 
quiescence light curves. These measurements lead to corresponding
disc radii of $R_d/R_\mathrm{L1}= 0.79\pm 0.08$, $0.62\pm 0.07$, and
$0.45\pm 0.06$, in very good agreement with the values derived
from the entropy landscape procedure. The measured ingress/egress
eclipse phases are depicted as pairs of vertical tick marks in
the left-hand panels of Fig.\,\ref{fig4}.

We used the expression of \cite{hh90} to compute a circularization
radius (the smallest possible radius for an inviscid accretion disc)
of $R_\mathrm{circ}/R_\mathrm{L1}= 0.0859\,q^{-0.426}(R_\mathrm{L1}/a)^{-1}=
0.30$. Thus, the measured quiescent disc radius of V4140 Sgr is
larger than its circularization radius at a $3\,\sigma$ confidence
level.

\subsubsection{Evolution of disc surface brightness} \label{evol}

\begin{figure*}
\includegraphics[scale=0.65,angle=270]{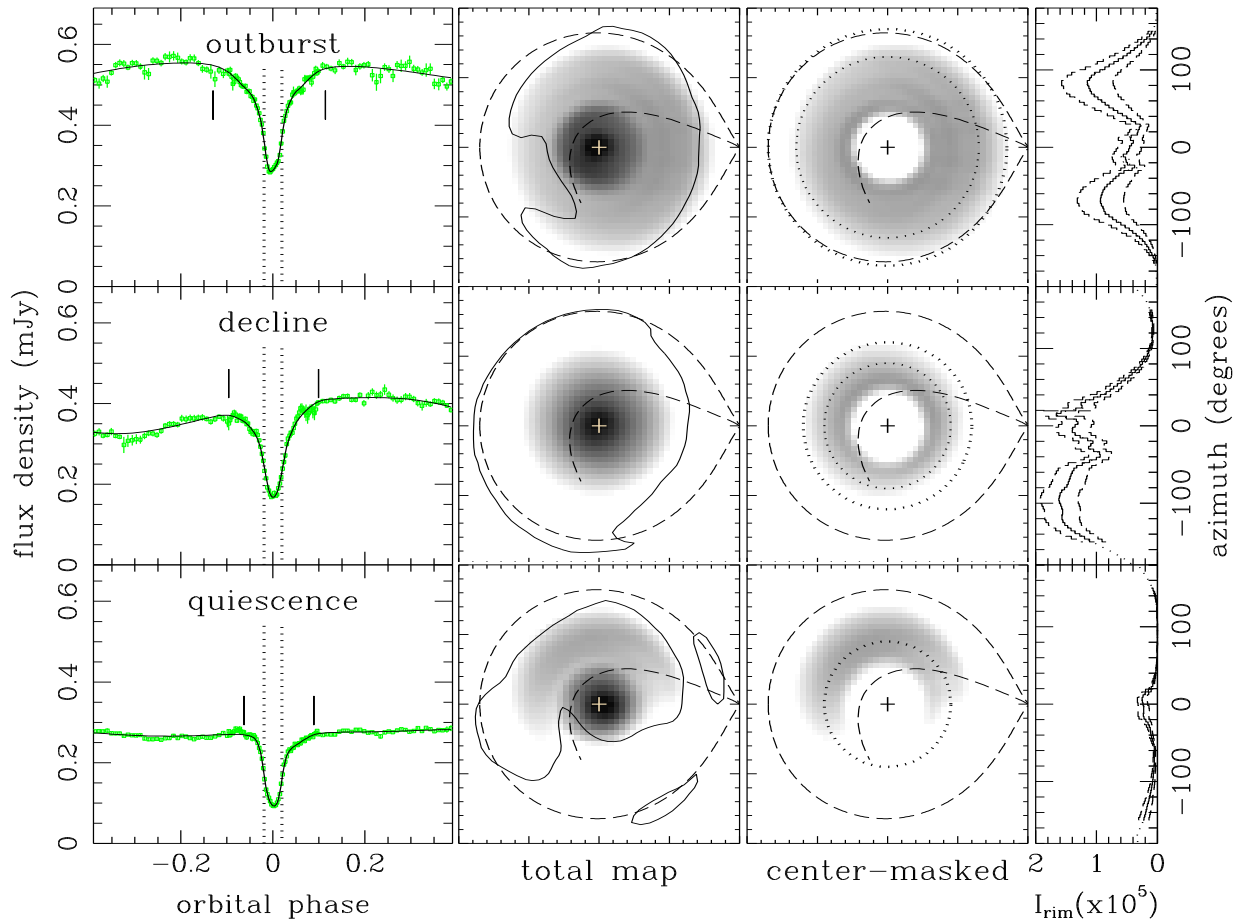}
\caption{ Left-hand: Data (points with error bars) and model (solid 
 lines) light curves for the outburst, decline and quiescence states.
 Vertical dotted lines mark the ingress/egress phases of the WD,
 while vertical tick marks depict the phases of eclipse 
 ingress/egress (see text). Middle-left: the corresponding eclipse 
 maps in a logarithmic grayscale (dark regions are brighter). A
 cross marks the position of the WD at disc centre. Dashed lines
 depict the primary Roche lobe and the gas stream trajectory.
 A contour line for S/N= 3 is overplotted on each map; features
 in the eclipse maps are statistically significant at or above the 
 $3\,\sigma$ level. The stars rotate counter-clockwise. 
 Middle-right: the eclipse maps with their central regions ($R< 
 0.35\,R_\mathrm{L1}$) masked for a clearer view of the brightness 
 distribution at the outer disc regions. Dotted circles mark the 
 minimum radius that matches the asymmetric outer disc brightness 
 distribution (for the outburst and decline maps) and the best-fit 
 disc rim radius. Right-hand: the intensity distribution
 along the disc rim. Dashed lines show the $1\,\sigma$ limit on the
 intensity distribution. Azimuths are counted from the line joining
 both stars and increase counter-clockwise.
}
\label{fig4}
\end{figure*}

Eclipse maps corresponding to the best-fit ($R_d,\beta$) values
are shown in Fig.\,\ref{fig4}, together with the data and model
light curves. The orientation of the eclipse maps is such that
the secondary star rotates counter-clockwise around the WD at
the centre of the map (in the observer's reference frame), while
the observer rotates clockwise (in the binary reference frame).
At phase zero, the secondary and the observer are on the right
side of the map (with the projected shadow of the secondary star
covering most of the disc). At phase $+0.25$~cycle the observer
is at the bottom (in the binary frame) while the secondary star
is at the top of the eclipse map (in the observer's frame).
%

The inner disc regions remain roughly at the same brightness level
through outburst, while significant brightness changes take place
in the outer disc regions. In order to emphasize these changes and
the brightness asymmetries in the outer disc regions, the inner disc
regions of the eclipse maps ($R< 0.35\,R_\mathrm{L1}$) were masked in
the middle-right panels of Fig.\,\ref{fig4}.

At outburst, the surface brightness distribution of the outer disc
is skewed towards the L1 point indicating that the disc is elliptical
at this stage. Two dotted circles are overplotted on the eclipse map
to help visualize the elliptical disc brightness distribution. The
outer circle corresponds to the disc rim $R_\mathrm{max}= R_d = 0.85\,
R_\mathrm{L1}$, while the inner circle has $R_\mathrm{min}= 0.65\,
R_\mathrm{L1}$. From these radii we estimate a disc eccentricity 
$e= (R_\mathrm{max} - R_\mathrm{min})/(R_\mathrm{max} + R_\mathrm{min})=
0.13$. For our outburst dataset the phase of superhump maximum 
coincides with mid-eclipse, right when the bulge of the elliptical
disc is aligned with the secondary star. 
Although less conspicuous, a similar brightness asymmetry is also
present in the decline eclipse map. In this case the major axis of
the ellipse is oriented at an angle $\simeq 100\degr$ with respect
to the line joining both stars, implying that the alignment of the
secondary star with its bulge occurs at binary phase $\simeq 
+0.28$~cycle. This is in good agreement with the observed phase
of superhump maximum in the decline light curve, $\phi_\mathrm{max}= 
+0.25$~cycle. From this eclipse map we find $R_\mathrm{max}= 
R_d= 0.6\,R_\mathrm{L1}$ and $R_\mathrm{min}=0.45\,R_\mathrm{L1}$, 
also leading to a disc eccentricity $e=0.13$.

These results are in line with the tidal resonance instability model,
which explains superhumps in terms of enhanced tidal dissipation 
(and extra disc emission) as the secondary star passes closer to 
the bulge of a slowly precessing elliptical disc 
\citep{whitehurst88,ho90,lubow91,lubow94}. For V4140~Sgr, the 3:1
resonance radius is $R_{31}= 0.66\,R_\mathrm{L1}$. At outburst, the
disc is comfortably larger than $R_{31}$ ($R_d=0.85\,R_\mathrm{L1}$),
whereas in the decline dataset the disc radius is slightly smaller
than $R_{31}$ ($R_d=0.6\,R_\mathrm{L1}$). 
The measured disc eccentricities are in good agreement with
the $e\simeq 0.1$ value found in numerical simulations of the
tidal resonance instability by \cite{hom91}.

Since the timescale of disc precession \citep[$\simeq$ 2-3 days,
see Table\,3.3 of][]{w95} is much longer than $P_\mathrm{orb}$, the
change in orientation of the elliptical disc along the time span
of the outburst and decline datasets ($\simeq 3$~h) is
$\Delta\theta \leq 20\degr$, smaller than the azimuthal smearing
effect adopted for the maximum entropy reconstructions
($\Delta\theta= 30\degr$). Therefore, there is no significant
azimuthal `blurring' of the outburst and decline eclipse maps
caused by the slow precession of the elliptical disc.

The right-hand panels of Fig.\,\ref{fig4} show the brightness
distributions along the disc rim. A word of caution is on
demand before we attempt to interpret these distributions.
One of the basic assumptions of the eclipse mapping method is
that the intensities in the eclipse map are independent of binary
phase, and that all changes in the light curve are caused by 
variable visibility and/or aspect of pixels with phase.
Superhumps clearly violate this assumption if they correspond 
to true, physical changes in intensity of accretion disc regions 
with binary phase. Therefore, the disc rim intensity distributions
in outburst and decline should not be interpreted as if there
is enhanced emission from extended regions at disc rim. Instead,
they just signal the azimuths (or binary phases) where enhanced
emission from the superhump light source occurs in each light curve.
Thus, the disc rim distributions tell us that superhump emission
is produced over a large fraction of the orbital period and is
centreed at phase $\phi_\mathrm{max}= 0\,\,\mathrm{and}\,+0.25$
in outburst and decline, respectively (of course, this can also be
inferred by direct inspection of the corresponding light curves).
However, an additional useful information is that, aside of the
elliptical disc bulge (responsible for the disc rim emission at
azimuth $\theta_\mathrm{sh}\simeq 0$), there are two other superhump
light sources in outburst, at azimuths $\theta_\mathrm{sh}\simeq
\pm (80\degr-90\degr)$ -- when the secondary star is roughly
orthogonal to the major axis of the elliptical disc. A similar
inference can be drawn from the decline disc rim distribution,
although in this case the two sources are closer in azimuth than
in outburst and the disc bulge contribution seems considerably
reduced or blended in azimuth with that from the two side sources.
These results are reminiscent to those of \cite{odonoghue}.
He applied a modified eclipse mapping technique to investigate
the superhump light source in Z~Cha when superhump maximum is
centreed in eclipse, and found that the extra emission arises
from three sources in the outer disc regions, one towards the
L1 point and two sources symmetrically located in the orthogonal
directions. These orientations coincide with the predicted 
positions where the elliptical orbits intersect \citep{hellier}.

Let us turn our attention to the quiescence light curve. 
Its eclipse map is dominated by emission from an extended 
asymmetric source with no distinguished contribution from the WD
(i.e., no sharp ingress/egress features in eclipse shape), 
indicating that efficient accretion through a high-viscosity disc
is taking place in quiescence. The long egress shoulder in the 
eclipse shape translates into enhanced emission along the outer 
disc regions ahead of the stream-disc impact point, suggesting
inefficient downstream cooling in the disc flow \citep{smak71}. 
On the other hand, the weak orbital hump maps into a disc rim 
narrow spot ($\Delta \theta= 40\degr$) at zero azimuth. 
The apparent hump phase of maximum $\phi_\mathrm{max}=
-0.08$~cycle is an illusion, as the spot starts being eclipsed
together with the accretion disc soon after that phase and is
behind the shadow of the secondary star when seen face on.
The azimuth of maximum emission of this spot
($\theta_\mathrm{max}=0\degr$) does not coincide
with the radial direction of the stream-disc impact point 
($\theta=31\degr$), but with the direction of the infalling gas
stream at that point. This is in contrast with results from the
dwarf novae U\,Gem \citep{marsh88}, OY\,Car \citep{wood89} and 
IP\,Peg \citep{tiago07}, where the maximum emission of the 
quiescent bright spot lies in the direction normal to a plane
roughly midway between the disc and stream flows.

\subsubsection{Radial temperature distributions} \label{radtemp}

In the DIM framework, a dwarf nova accretion disc cycles between 
a hot, high-viscosity state where hydrogen is fully ionized 
(outburst) and a faint, low-viscosity state where hydrogen is
mostly neutral (quiescence) \citep{lasota,acpower}. The 
instability that sets this limit-cycle behavior is tied to 
the partial ionization of hydrogen -- which leads to tight 
constraints in disc gas temperature along the outburst cycle. 
Namely, there is a critical effective temperature $T_\mathrm{crit2}$
above which the disc gas should stay while in outburst,
and a critical effective temperature $T_\mathrm{crit1}$ below
which it should remain while in quiescence \citep{lasota},
\begin{equation}
 T_\mathrm{crit1}= 5800\,\left(\frac{M_1}{M_\odot}\right)^{0.03}
    \left( \frac{R}{10^{10}\,cm} \right)^{-0.09} \,K \,,
\label{eq-tcrit1}
\end{equation}
\begin{equation}
 T_\mathrm{crit2}= 7200\,\left(\frac{M_1}{M_\odot}\right)^{0.03}
    \left( \frac{R}{10^{10}\,cm} \right)^{-0.08} \,K \,.
\label{eq-tcrit2}
\end{equation}
Decline from outburst is understood in terms of the inwards
propagation of a cooling front from the outer disc regions, 
progressively transitioning the disc from its hot state back to 
the cool quiescent state. Disc regions ahead of the cooling front 
should still have $T>T_\mathrm{crit2}$, while those behind the 
cooling front should show $T<T_\mathrm{crit1}$.

This prediction can be tested with eclipse mapping.
If the distance to the binary is known, the usual practice
is to convert the intensities in the eclipse maps to blackbody
brightness temperatures $T_b$, which can then be compared to
$T_\mathrm{crit}$ and to the $T_\mathrm{eff}(R) \propto R^{-3/4}$
law of opaque, steady-state disc models. A valid criticism about
this procedure is that $T_b$ may not be a proper estimate of the
disc effective temperature $T_\mathrm{eff}$.
As pointed out by \cite{bal98}, a relation between these two
quantities is non-trivial, and can only be properly obtained by
constructing self-consistent models of the vertical structure of
the disc. On the other hand, observations indicate that the colors
of optically thick disc regions are in between those of blackbody
and main-sequence stars \citep{hc85,hs85,bbb96,bsh96} -- in which
cases a broad-band, blackbody brightness temperature becomes a
fair and useful approximation to $T_\mathrm{eff}$.
\cite{bb05} applied a method similar to cluster main-sequence
fitting to an accretion disc color-magnitude diagram of V4140\,Sgr
to infer a distance of $170\pm 30$\,pc. Their eclipse maps lead
to radial brightness temperatures in outburst and quiescence which
are consistent at the 1\,$\sigma$ limit in the $B$, $V$ and $R$
passbands, suggesting that the accretion disc is indeed optically
thick and that $T_b \simeq T_\mathrm{eff}$ is a reasonable
approximation in this case.

\begin{figure}
\includegraphics[scale=0.45]{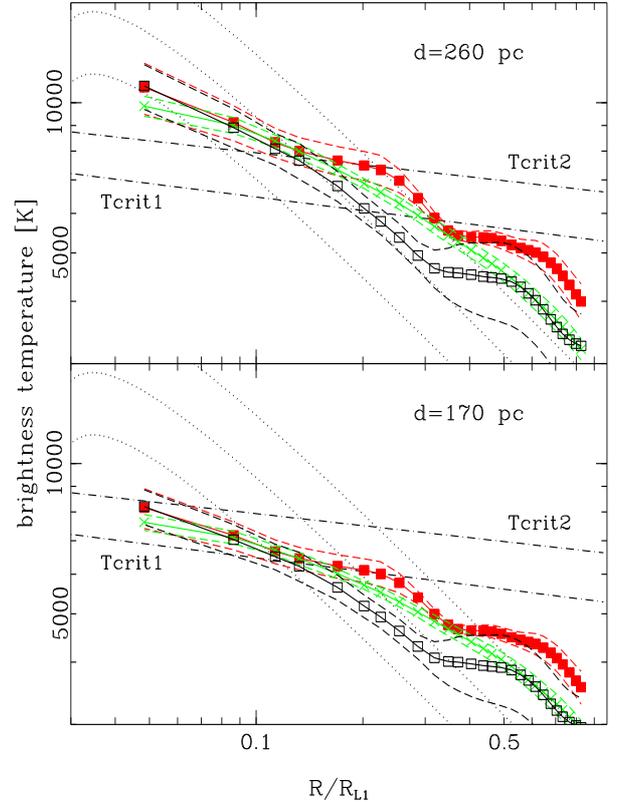}
\caption{ Azimuthally-averaged radial brightness temperature 
 distributions for the outburst (solid squares, red), decline 
 (crosses, green) and quiescence (open squares, black) accretion
 disc maps, for radial bins of width $0.03\,R_\mathrm{L1}$. 
 Dashed lines show the 1\,$\sigma$ limits on the
 average temperatures. Dotted lines correspond to steady-state
 optically thick disc models for mass accretion rates of $10^{-10}, 
 10^{-10.5}\, \mathrm{and}\, 10^{-11}\,M_\odot\,\mathrm{yr}^{-1}$, 
 respectively from top to bottom. Dot-dashed lines show the 
 critical effective temperatures $T_\mathrm{crit2}$ (above which
 the disc gas should stay while in outburst) and $T_\mathrm{crit1}$
 (below which it should remain while in quiescence), according
 to the DIM \citep[e.g.,][]{w95}. The distributions are shown for
 assumed distances of 260\,pc (top) and 170\,pc (bottom).
}
\label{fig5}
\end{figure}

Fig.\,\ref{fig5} shows azimuthally-averaged radial brightness 
temperature distributions for the outburst (solid squares), 
decline (crosses) and quiescence (open squares) accretion disc maps
for assumed distances of 170\,pc (bottom) and 260\,pc \citep[top,
the upper 3\,$\sigma$ limit on the distance estimate of][]{bb05}.
The intensities in the eclipse maps were scaled to the assumed
distances and corrected for interstellar extinction effects
before being converted to brightness temperatures.
From the galactic interstellar contour maps of \cite{lucke78},
we estimate a reddening of $E(B-V)= 0.4\;\mathrm{mag\,kpc}^{-1}$
towards the direction of V4140\,Sgr.
%

The temperatures in the inner disc regions ($R<0.1\,R_\mathrm{L1}$)
are consistently the same throughout outburst. The outer 
disc regions ($R > 0.1\,R_\mathrm{L1}$) progressively cool from 
outburst to quiescence.
At a distance of 170\,pc the whole outburst cycle proceeds at 
temperatures below $T_\mathrm{crit2}$. Increasing the distance to 
the binary increases the intrinsic intensities and corresponding
disc brightness temperatures. But even at a distance of 260\,pc
most disc regions ahead of the inwards moving cooling wave show
temperatures below $T_\mathrm{crit2}$.
In order to reconcile the observations with DIM, one needs to
increase the distance above 300\,pc. However, this steepens the
radial temperature distribution in quiescence, bringing it into
good agreement with the $T\propto R^{-3/4}$ law of high-viscosity,
opaque steady-state discs -- in contradiction to the DIM
expectation that quiescent disc of dwarf novae should have low
viscosities ($\alpha \sim 0.01$), long viscous timescales
and, therefore, would hardly have time to reach a steady-state 
in between consecutive outbursts.
Furthermore, this also increases the disc temperatures in
quiescence, bringing an increasing (and significant) fraction of
the quiescent disc into contradiction with the DIM requirement
$T < T_\mathrm{crit1}$. In other words, the disc temperatures do
not differ sufficiently in outburst and quiescence that one can
have both $T < T_\mathrm{crit1}$ everywhere in quiescence and
$T > T_\mathrm{crit2}$ over most of the disc in outburst.
It seems clear that {\em it is not possible to reconcile the
outbursts of V4140~Sgr with DIM}.

\subsection{Flickering mapping in quiescence} \label{quies}

\begin{figure}
\includegraphics[scale=0.35,angle=270]{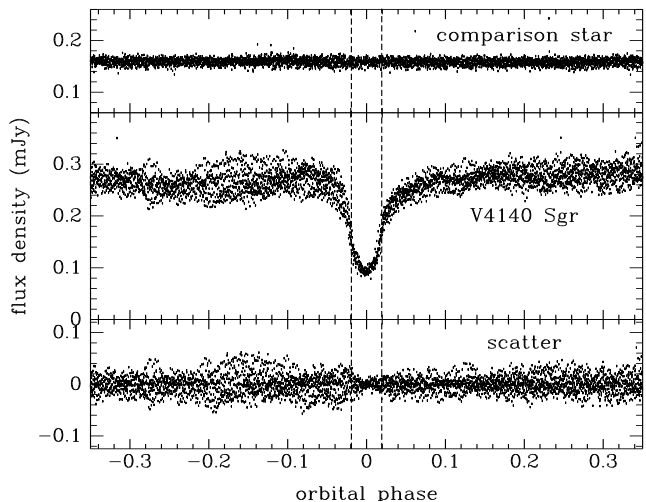}
\caption{ The quiescent light curves of V4140\,Sgr (middle) and 
of a comparison star of similar brightness (top). The lower 
panel shows the light curves of the middle panel after subtraction
of the average orbital curve; the scatter around the mean flux
yields an indication of the flickering amplitude at each phase.
Vertical dashed lines mark the ingress/egress phases of the WD. }
\label{fig6}
\end{figure}

Figure\,\ref{fig6} shows the individual quiescent light curves
of V4140~Sgr superimposed in phase.
%
%
Along the $\simeq 50$\,d time interval covered by the
quiescence observations, V4140~Sgr showed no systematic change or
long-term trend in eclipse depth, out-of-eclipse brightness or
eclipse width. Therefore, the light curves of that period can be
assigned to a single brightness state, and all observed
variability can be attributed to flickering.
The top panel depicts light curves of a comparison star of similar
brightness. The scatter with respect to the mean flux is perceptibly
larger in V4140\,Sgr than in the comparison star and is caused by 
flickering. The scatter is larger close to orbital hump maximum
(suggesting that the bright spot contributes to the flickering)
and is smaller during eclipse (indicating that flickering sources
are occulted at these phases).

We applied the ``single'' \citep{bruch96,bruch00} and ``ensemble''
\citep{hs85,bennie} techniques to the quiescent light curves of
V4140\,Sgr to derive the orbital dependency of its steady-light 
and of the low- and high-frequency flickering components. The
reader is referred to \cite{bb04} for a detailed description and
combined application of both techniques.
While the ensemble technique samples flickering at all frequencies,
the typical power law dependency of the flickering in CVs implies 
that an ensemble curve is dominated by the low-frequency flickering
components. On the other hand, the high-pass filtering process
of the single technique leads to curves which sample the 
high-frequency flickering. The combination of both methods allows
one to separate the low- (ensemble) and high-frequency (single)
components of the flickering sources.

Because all light curves are at the same brightness level, 
applying the ensemble technique becomes a matter of computing the 
mean flux at each phase bin (the steady-light curve) and the
standard deviation with respect to the mean (the scatter curve,
$\sigma_\mathrm{tot}(\phi)$, with contributions from the Poisson
noise, $\sigma_\mathrm{Poi}(\phi)$, and from the flickering,
$\sigma_\mathrm{flick}(\phi)$). The angular coefficient (which
measures the long-term brightness changes) is consistently zero
within the uncertainties. The ensemble flickering curve is
obtained by correcting the scatter curve from the Poisson noise
contribution,
\begin{equation}
  \sigma_\mathrm{flick}(\phi)= \sqrt{\sigma_\mathrm{tot}^2(\phi) -
    \sigma_\mathrm{Poi}^2(\phi) }\,\, .
  \label{eq-flick}
\end{equation}

\begin{figure}
\includegraphics[scale=0.34,angle=270]{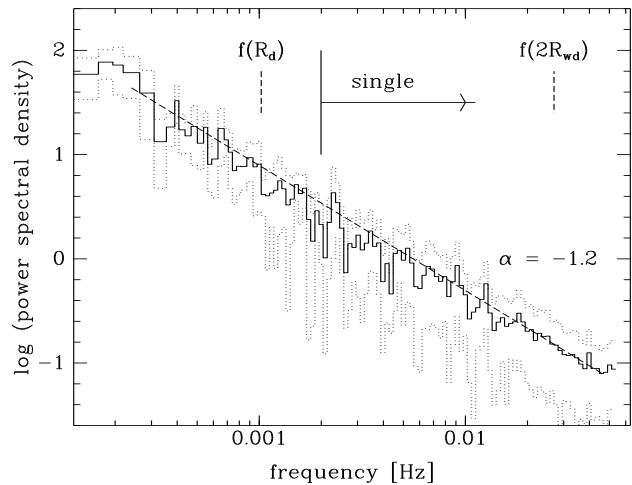}
\caption{ Average PDS. Dotted lines show the 1~$\sigma$ limits
on the average power. The best-fit power law $P(f)\propto f^{-1.2}$
is shown as a dashed line. A vertical tick marks the low-frequency
cut-off of the filtering process applied to derive the single 
scatter curve. The Keplerian frequencies corresponding to $2\,R_{wd}$
and to the quiescent disc radius $R_d$ are illustrated as vertical
dashed ticks. }
\label{fig7}
\end{figure}

The average steady-light curve was subtracted from each individual
light curve to remove the DC component, and a Lomb-Scargle 
periodgram \citep{numrec} was calculated. The periodgrams of
all light curves were combined to yield a mean periodgram and
a standard deviation with respect to the mean. Figure\,\ref{fig7}
shows the resulting average power density spectrum (PDS)
binned to a resolution of 0.02 units in log(frequency).
%
%
The PDS is well described by a power law $P(f)\propto f^{-1.2}$,
which becomes flat for $f_\mathrm{low}< 0.2$\,mHz ($t_\mathrm{low}>83$
minutes) and disappears in the white noise for $f_\mathrm{high}>40$
\,mHz ($t_\mathrm{high}<25$\,s). The slope of the PDS distribution
is consistent with the general trend found in other CVs, which
can be well described by power laws with average exponent
$P(f)\propto f^{-2.0 \pm 0.8}$ \citep{bruch92,bb04,bb08}.

We followed \cite{bb08} and subtracted the average
steady-light curve from each individual light curve before applying
the single filtering process in order to eliminate the steep
gradients produced by the eclipse in the light curve. Our single
light curve includes flickering components with frequencies
$f_c> 2$\,mHz (timescales shorter than 500\,s). Single curves
obtained with higher cutoff frequencies show the same morphology but
are noisier (because of the reduced power) and lead to less reliable
results, while for those obtained with lower cutoff frequencies 
the excess of filtering starts smearing out the eclipse of the
flickering. The single flickering curve is obtained by correcting
the corresponding scatter curve from the Poisson noise contribution
(cf. Eq.\,\ref{eq-flick}).

As the disc opening angle in quiescence is negligible 
($\beta= 0.5\degr$, Sect.\,\ref{discrim}), the steady-light and
the flickering curves were analyzed by a standard eclipse mapping
algorithm \citep{bs93}, with a flat Cartesian grid of $51\times 51$ 
pixels centreed on the WD and side $2\,R_\mathrm{L1}$ in the orbital
plane, to solve for a map of the disc surface brightness distribution
in each case.
Because this version of the eclipse mapping method does not take 
into account out-of-eclipse brightness changes, these were removed
from the light curves by fitting a spline function to the phases
outside eclipse, dividing the light curve by the fitted spline, 
and scaling the result to the spline function value at phase zero.

\begin{figure*}
\includegraphics[scale=0.63,angle=270]{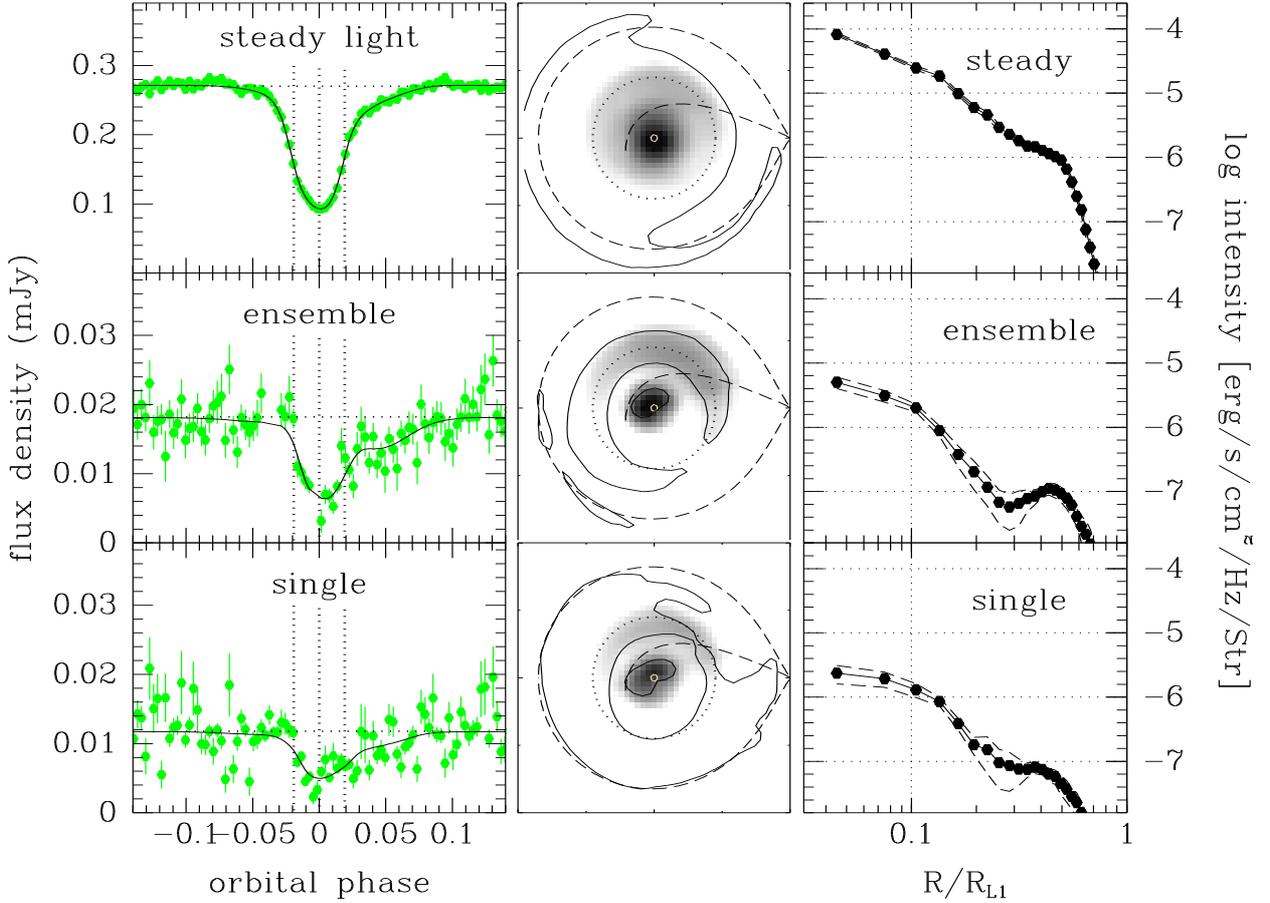}
\caption{ Left: Data (points with error bars) and model (solid lines)
 light curves for the steady-light, the ensemble and the single 
 flickering curves. Vertical dotted lines mark the ingress/egress
 phases of the WD and mid-eclipse. Middle: the corresponding eclipse
 maps in a logarithmic grayscale (dark regions are brighter).  A
 small circle marks the position and size of the WD at disc centre.
 The notation is the same as in Fig.\,\ref{fig4}. Features in the
 eclipse maps are statistically significant at or above the
 $3\,\sigma$ level.
 Right: Azimuthally-averaged radial intensity distributions of
 the eclipse maps in the middle panels. Dashed lines show 
 the $1\,\sigma$ limits on the average intensities. }
\label{fig8}
\end{figure*}

The adopted eclipse geometry ($\Delta\phi, q, i$) and default map
parameters ($\Delta r, \Delta\theta$) are the same as in
Sect.\,\ref{erup}, and the reconstructions also reached a final
reduced chi-square $\chi_\mathrm{red}^{2}\simeq 1$ in all cases.
Monte Carlo simulations with the bootstrap technique were similarly
performed to obtain the uncertainties in the eclipse maps. These
are used to draw the contour maps of Fig.\,\ref{fig8}, and to
estimate the uncertainties in the radial distributions shown in
Fig.\,\ref{fig9}.

The resulting steady-light, ensemble and single curves have phase
resolution of 0.003\,cycle and cover the phase range ($-0.15,+0.15$)
cycle. These light curves and corresponding eclipse maps are
shown in Fig.\,\ref{fig8}.
%

\subsubsection{The steady-light} \label{steady}

The steady-light curve is the quiescence light curve of
Sect.\,\ref{erup} after removal of the weak orbital hump.
Consistently, the steady-light eclipse map is very similar to
the quiescence eclipse map (Fig.\,\ref{fig4}), showing a broad
brightness distribution of an asymmetric accretion disc which
extends up to $0.45\,R_\mathrm{L1}$. This brightness distribution
was already discussed in Sect.\,\ref{evol}.

It is worth noting that, although derived from essentially the
same light curve, the steady-light map is of higher statistical
significance than its cousin quiescence map. This reflects a
subtle but relevant difference between the standard and the 3D
eclipse mapping. Standard eclipse mapping is simpler and more
robust to noise in the light curve (e.g., flickering) as only a
relatively narrow phase range of the light curve covering the
eclipse shape is used to derive the disc brightness distribution.
Because 3D eclipse mapping uses the whole orbital light curve to
derive the disc + rim brightness distributions, it is more
sensitive to noise in the light curve and any flare/flicker
outside of eclipse contributes to the uncertainty of all visible
pixels at that phase. Nevertheless, the intrinsic lower statistical
significance of the 3D eclipse mapping seems a reasonable price
to pay in return for the extra information about the disc radius
and half-opening angle delivered by this technique.
On the other hand, the higher statistical significance with which
maps are obtained with the standard eclipse mapping is key to
allow statistically meaningful results in the derivation of the
radial distribution of the flickering amplitude (Sect.\,\ref{flick}).

\subsubsection{low- and high-frequency flickering} \label{flick}

The ensemble and single flickering curves have similar shapes,
showing the deeper eclipse of a central source plus an extended
egress shoulder. This eclipse shape maps into an asymmetric source
along the disc rim ahead of the stream-disc impact point ($R\simeq
0.45\,R_\mathrm{L1}$, stream-disc impact flickering) and an extended
central source several times larger in radius than the WD at disc
centre (boundary layer + disc-related flickering)
\footnote{A small circle at the centre of each eclipse map depicts
  the size of the WD to allow a direct comparison with the radial
  extent of the disc-related flickering component.}.
The solid contour line overplotted on each eclipse map depicts
the $3\,\sigma$ confidence level region as derived from the map
of statistical significance in each case. The two flickering
sources are statistically significant at or above the $3\,\sigma$
confidence level both in the ensemble and in the single map.

The disc-related flickering component is not centred at the WD
position, but is slightly skewed towards the ballistic stream
trajectory. We tested whether this was an artifact caused by an
error in the choice of phase zero by obtaining eclipse maps for
versions of the light curves with small offsets in phase and by
allowing changes in the adopted eclipse geometry within the
uncertainties in the binary parameters $q$ and $i$. These
simulations indicate that it is not possible to simultaneously
centre the steady-light brightness distribution and the disc
flickering component, and we conclude that the slight offset of
the disc flickering component is real.

Azimuthally-averaged radial intensity distributions for the
steady-light, ensemble and single flickering maps are shown in the
right-hand panels of Fig.\,\ref{fig8}. Each eclipse map is separated
in radial bins of width $0.03\,R_\mathrm{L1}$ and the median intensity
is computed for each bin. The associated uncertainty is obtained
by computing the median absolute deviation with respect to the
median intensity for the set of bootstrap reconstructions of the
corresponding map and multiplying the result by 1.48 to find the
standard deviation \citep[e.g.,][]{stat}.

\begin{figure}
\includegraphics[scale=0.46]{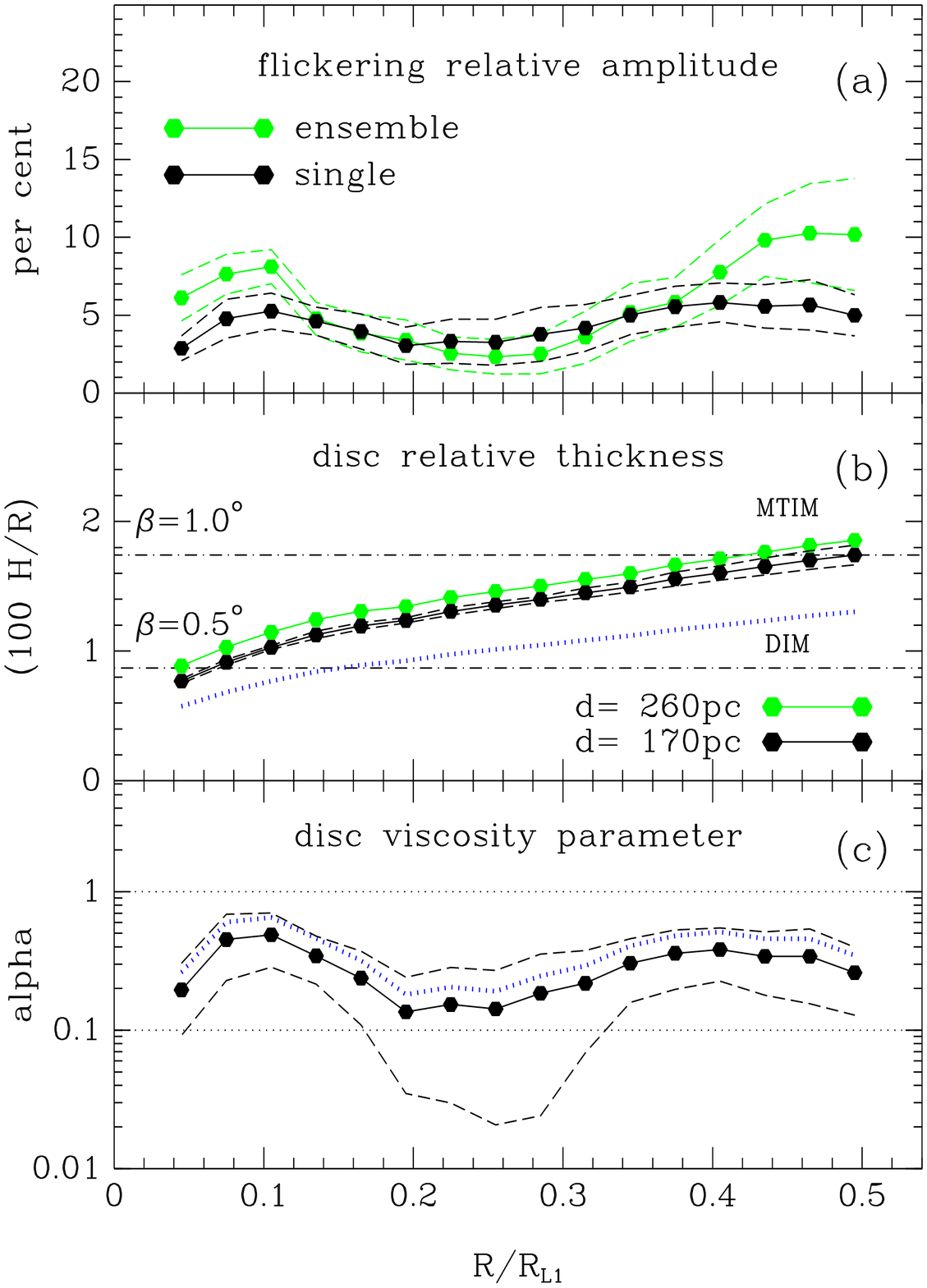}
\caption{ Azimuthally-averaged radial distributions. (a) The
  relative amplitude of the ensemble and single flickering. Dashed
  lines show the $1\,\sigma$ limits on the average values. (b)
  The disc relative thickness ($H/R$) derived from the $T_b(R)$
  distribution assuming $s=2$ (MTIM case) for distances of 170\,pc
  (black) and 260\,pc (green). Dashed lines show the $1\,\sigma$
  limits on the $H/R$ distribution derived from the uncertainties
  in the $T_b(R)$ distribution. The dotted line corresponds to
  the $H/R$ distribution assuming $s=1.5$ (DIM case).
  Horizontal dot-dashed lines depict the $H/R$ value
  derived from the measured disc half-opening angle
  (Sect.\,\ref{discrim}) and its $1\,\sigma$ upper limit. (c) The
  disc viscosity parameter $\alpha$ derived using the Gertseema \&
  Achterberg (1992) model. The solid line corresponds to the H/R
  distribution for the MTIM case ($s=2$); dashed lines show the
  $1\,\sigma$ limits on the derived $\alpha$ value. A dotted line
  depicts the corresponding distribution for the DIM case ($s=1.5$).
}
\label{fig9}
\end{figure}

Fig.\,\ref{fig9}a compares the relative amplitude of the ensemble
and single flickering in V4140~Sgr. The radial run of the relative
amplitudes are obtained by dividing the radial intensity distribution
of these two flickering maps by that of the steady-light. Dashed
lines show the $1\,\sigma$ limits on the average flickering
amplitudes. The distributions are not reliable for
$R > 0.5\,R_\mathrm{L1}$ because of the reduced statistical
significance of the flickering maps and the rapidly declining
intensities in the steady-light map, and are not shown in
Fig.\,\ref{fig9}a.
%

The difference between the ensemble and single distributions tells
us where the low-frequency flickering comes from. An excess of
ensemble over single flickering at $R\leq 0.12\,R_\mathrm{L1} \simeq
5\,R_\mathrm{wd}$ indicate the existence of a low-frequency flickering
central source at the $2.5\,\sigma$ confidence level.
The radial extent of this source is too large to be accounted for
by the radial smearing effect of the entropy on the eclipse map
\footnote{In our eclipse maps, point sources are smeared in the radial
  direction by $3\,\Delta r= 0.06\,R_\mathrm{L1}\simeq 2.4\,R_\mathrm{wd}$.}
\cite[see][]{bsh96}, indicating that the source is larger than the
WD by a factor $\simeq 2-3$. This central flickering component may
possibly arise from optical reprocessing of intermitent x-ray and
UV irradiation by the boundary layer at the atmosphere of the
innermost disc regions.
Additional excess of ensemble over single flickering is seen in
the outer disc regions ($R > 0.4\,R_\mathrm{L1}$, at the
$3\,\sigma$ confidence level), which is readily associated to
the stream-disc impact region. This stream-disc flickering
component is relatively stronger ($\simeq 5$ per cent amplitude)
than the central source flickering ($\simeq 3$ per cent
amplitude). There is no low-frequency flickering produced in the
intermediate disc regions ($0.12 < R/R_\mathrm{L1} < 0.4$).
The high-frequency flickering is spread over the accretion disc,
with a $\simeq 4-5$ per cent amplitude roughly independent of disc
radius. This is the disc-related flickering component. The slight
increase in high-frequency flickering amplitude at the innermost
and outer disc regions might be due to some high-frequency
contribution from the central and stream-disc flickering components.

Our ensemble and single flickering maps allow us to disentangle
three different sources of flickering in V4140\,Sgr.
The presence of a central flickering and of a stream-disc impact
flickering sources are in good agreement with the scenario put
forward by \cite{bruch92,bruch00}. If the central flickering
component arises in the boundary layer its spectrum may be much
bluer than the inner disc spectrum, and our optical flickering
mapping experiment is underestimating its contribution to the
total flickering. In this case, this boundary layer flickering
should dominate the UV light curve of V4140\,Sgr, with fast
reduction in flickering activity as the WD goes into eclipse.
Our flickering mapping experiment also reveals the existence of
a disc flickering component in V4140\,Sgr, which allow us to probe
the viscosity in its quiescent disc (see Sect.\,\ref{discuss}).
This disc flickering component accounts for 2/3 of the total
optical flickering in V4140\,Sgr. The results bear some
resemblance with those from the quiescent dwarf nova V2051\,Oph,
where the low-frequency flickering is associated with an
overflowing gas stream while the high-frequency flickering
arises from the accretion disc \citep{bb04}.

\section{The quiescent disc viscosity} \label{discuss}

The identification of a disc-related flickering component enables us
to follow the steps of \cite{bb04} and to use the model of \cite{ga}
to estimate the magnitude of the viscosity parameter $\alpha$ for
the quiescent disc of V4140\,Sgr. This allows us to critically test
the largely different predictions of DIM and MTIM for a quiescent
disc of a dwarf nova. As the expression relating $\alpha$ and
$\sigma_\mathrm{D}/\langle D \rangle$ involves the disc relative
thickness $H/R$ and because this quantity is loosely constrained by
the entropy landscape analysis, we start our exercise by using the
measured radial temperature distribution for an independent (and
more precise) measurement of $H/R$ as well as to check the
consistency of the thin disc result of Sect.\,\ref{discrim}.

In the atmosphere of an opaque, $\alpha$-disc model,
\begin{equation}
  {T_c}^{1/2} = s\, {T_\mathrm{eff}}^{1/2} \simeq s \, {T_b}^{1/2} \, ,
\label{eq:temp}
\end{equation}
where $T_c$ is the mid-plane disc temperature, $s$ is the square root
of the mid-plane to the effective temperature ratio and the final
step is justified by the discussion in Sect.\,\ref{radtemp}.
An estimate of the $s$ value is obtained by comparing the
analytical fits of the critical effective temperatures
($T_\mathrm{crit1},T_\mathrm{crit2}$) and of $T_c$ \citep[see, e.g.,
eqs.\ A5-A8 in][]{lasota}; for the upper turning point ($\alpha
\sim 0.1$), $s= (T_c/T_\mathrm{crit2})^{1/2} \simeq 2$, while for the
lower turning point ($\alpha \sim 0.01$),
$s= (T_c/T_\mathrm{crit1})^{1/2} \simeq 1.5$.
\cite{hirose14} computed detailed equilibrium vertical atmospheres
of $\alpha$-discs subject to MHD turbulence for a set of physical
conditions of interest to dwarf nova accretion discs. Their model
for a high-viscosity ($\alpha\simeq 0.1$), hot and opaque disc with
$T_\mathrm{eff}=7490\,K$ (representative of the range of disc
temperatures found in V4140\,Sgr) consistently has $T_c=
4\,T_\mathrm{eff}$ (or $s=2$).
Thus, we adopt $s=2$ for the case of a high-viscosity quiescent
disc (as expected in the MTIM framework) and we take $s=1.5$ as
corresponding to a low-viscosity quiescent disc (as expected in
the DIM framework).

We may then use the inferred $T_b(R)$ distribution and the thin
disc approximation to estimate the radial run of the relative disc
thickness $H/R$ \citep{acpower},
\begin{equation}
  \frac{H}{R} = \frac{c_s}{v_k} =
  \left( \frac{k T_c}{\mu m_H} \frac{R}{GM_\mathrm{wd}} \right)^{1/2}
  \propto s\left[ R\,T_b(R) \right]^{1/2}  ,
 \label{eq-thick}
\end{equation}
where $k$ and $G$ are, respectively, the Boltzmann and the gravitation
constants, $\mu$ is the mean molecular weight, $m_H$ is the hydrogen
atom mass, and we adopt $\mu=0.615$ (adequate for a fully ionized
'cosmic' misture of gases), $M_\mathrm{wd}=0.73\pm 0.08 \,M_\odot$
\citep{bb05}. The disc relative thickness obtained with
Eq.(\ref{eq-thick}) assuming $s=2$ (MTIM case) is plotted in
Fig.\,\ref{fig9}b for distances of 170\,pc (black) and 260\,pc
(green). An $H/R$ distribution computed assuming $s=1.5$ (DIM case)
and distance of 170\,pc is shown as a dotted line in
Fig.\,\ref{fig9}b. The uncertainties in the $H/R$ distribution are
dominated by the uncertainty in the temperature scaling factor $s$.
For a given $T_\mathrm{eff}$ value, a quiescent DIM disc is thinner
than its MTIM counterpart because its mid-plane temperature is lower
and, therefore, $c_s$ and $H$ are also lower. The fact that the
$H/R$ distributions increase with radius imply that the disc face
is concave and, therefore, there is room for reprocessing of x-ray
and UV radiation from the boundary layer in the disc atmosphere
\cite[e.g.,][see Sect.\,\ref{flick}]{acpower}.
The disc relative thickness derived from the disc half-opening
angle ($\tan\beta= H/R= 0.0087\pm 0.0087$) is shown as horizontal
dot-dashed lines for comparison.
Disc relative thicknesses derived independently from the disc
temperatures and from the disc half-opening angle are consistent
with each other at the $1\,\sigma$ level, and indicate that the
quiescent disc of V4140\,Sgr is geometrically thin.

Once we have the radial runs of the relative amplitude of the disc
flickering and of the disc thickness, we may use the MHD turbulence
model of \cite{ga} to estimate the radial run of the viscosity
parameter in the quiescent disc of V4140\,Sgr. In their model, the
number of turbulent eddies that contribute to the local fluctuation
is,
\begin{equation}
  N(R) = 4\,\pi \frac{R}{H}\,\left( \frac{H}{L} \right)^2 \, ,
  \label{eq-ga1}
\end{equation}
where $L$ is the size of the largest turbulent eddy. The local rms
value of the fluctuations $\sigma_\mathrm{D}(R)$ in the average
energy dissipation rate per unit area $\langle D(R) \rangle$ is
found to be,
\begin{equation}
 \sigma_\mathrm{D}(R)\simeq 2.5\,\langle D(R)\rangle /\sqrt{N(R)}\,,
 \label{eq-ga2}
\end{equation}
while the disc viscosity parameter can be written as,
\begin{equation}
 \alpha \equiv \frac{\langle T_{r\phi} \rangle}{P} =
 \frac{3\,\nu_t}{2\,c_s\,H} \simeq 0.9\,\left( \frac{L}{H} \right)^2
 \, ,
 \label{eq-ga3}
\end{equation}
where $\langle T_{r\phi} \rangle$ is the local average shear stress
(Maxwell + Reynolds), $P$ is the local pressure, and $\nu_t$ is the
local (turbulent) disc viscosity. Combining
Eqs.\,(\ref{eq-ga1}-\ref{eq-ga3}), we obtain the radial run of the
disc viscosity parameter,
\begin{equation}
 \alpha(R) \simeq 0.46\, \left[ \frac{100\,H}{R} \right]^{-1} \left[
 \frac{\sigma_\mathrm{D}(R)}{0.05\,\langle D(R) \rangle} \right]^2 \, .
\label{eq-flick-ampl}
\end{equation}
If the disc-related flickering is caused by fluctuations of the energy
dissipation rate induced by MHD turbulence in an optically thick
disc, the relative amplitude of this disc flickering component yields
a good estimate of the ratio $\sigma_\mathrm{D}/\langle D \rangle$.

The resulting $\alpha(R)$ distributions are shown in Fig.\,\ref{fig9}c.
Because of the dependency $H/R \propto {T_b}^{1/2}$, the influence of
errors in distance and temperature on $\alpha$ is negligible.
Errors in $s$ also have a relatively small effect to the uncertainty
of $\alpha$ in comparison with the large contribution of the errors
in the flickering relative amplitude curves (because of the
dependency $\alpha\propto [\sigma_\mathrm{D}/\langle D \rangle]^2$
and the relatively lower accuracy of the latter quantity).
Large values of $\alpha\simeq 0.2-0.4$ at found all disc radii,
in good agreement with the MTIM expectations.
In order to model the long (80-90\,d) time interval between outbursts
of V4140\,Sgr in the DIM framework, an $\alpha_\mathrm{quies}\simeq
0.01$ is required \citep{c93,w95,lasota}. This corresponds to the
lower y-axis end of Fig.\,\ref{fig9}c. The derived $\alpha$'s are
systematically larger than the DIM expectation by at least an order
of magnitude. Outside the region $0.2\leq R/R_\mathrm{L1}\leq 0.3$,
the discrepancy is significant at the $2\,\sigma$ limit. Computing
$\alpha(R)$ from the $H/R$ distribution of the DIM case ($s=1.5$,
shown as a dotted line in Fig.\,\ref{fig9}c) only increases
the discrepancy, as the thinner DIM disc leads to even larger
$\alpha$ values. If the MHD turbulence model of \cite{ga} is
correct, the above results indicate that the quiescent accretion
disc of V4140\,Sgr has high-viscosity.

The V4140\,Sgr quiescent data allows an additional test of DIM,
which is independent of any MHD turbulence model. One of the
Achilles' heel of DIM is the prediction (unmatched by observations)
that the average brightness of a dwarf nova should increase
between successive outbursts as matter piles up in a low-viscosity
quiescent disc, steadily increasing the disc surface density,
temperature and, therefore, brightness \citep[e.g.,][]{lasota}.
In an attempt to overcome the discrepancy, \cite{truss} proposed
that the unobserved increase in disc brightness along quiescence
could be compensated for by the gradual cooling of a small, hot
and high-viscosity inner disc region. Their two-dimensional
time-dependent numerical models of dwarf nova accretion discs
seem to support this proposal. Could the observed high-viscosity
quiescent disc of V4140~Sgr be the gradually cooling hot inner
disc region proposed by \cite{truss}?

In an eclipsing dwarf nova, the gradual cooling of a hot inner
disc region combined with the simultaneous increase in brightness
of the outer regions plus the gradual contraction of a
low-viscosity, non-steady quiescent disc should make its eclipse
progressively narrower (because of disc shrinkage) and shallower
(because of the reduced brightness contrast between inner and outer
disc). There is no observational support for any of these effects
in our V4140~Sgr quiescent data. The shape, width and depth of the
quiescent eclipse is consistently the same within the uncertainties
along the $\simeq 50$\,d time interval covered by our quiescence
observations. This observational window corresponds to more than
50 per cent of the typical V4140~Sgr outburst recurrence time
\citep{bb05} and started right after the end of a superoutburst.
The stability of our observations not only do not support the
prediction of \cite{truss}, but they also indicate that the
quiescent accretion disc of V4140~Sgr is in a steady-state regime.
Again, this is in excellent agreement with the picture of a
high-viscosity quiescent accretion disc.

If the quiescent accretion disc of V4140~Sgr is already in the
high-viscosity regime, there is no room for disc instabilities to
set in and it is not possible to explain its outbursts in terms
of DIM. Moreover, it is not possible to reproduce its short (5-10\,d)
outburst length in comparison with the relatively long time interval
between sucessive outburst (80-90\,d) in the limit-cycle of the
DIM framework since there is no further room to increase $\alpha$
by an order of magnitude from quiescence to outburst.
In combination with the observed outburst disc temperatures
(below those required by DIM, $T_b < T_\mathrm{crit2}$,
Sect.\,\ref{radtemp}), these results strengthen the conclusion that
the outbursts of V4140~Sgr are driven by bursts of mass transfer
and not by disc instabilities \citep[c.f.,][]{bb05}.

V4140~Sgr is not the only dwarf nova showing observational evidence
for outbursts of mass transfer, but the younger member of an
increasing list including V2051~Oph \citep{bb04,bap07}, EX~Dra
and HT~Cas \citep{bc01,bap12}, V513~Cas and IW~And \citep{hl14},
and possibly EX~Hya \citep{hel2000}.

\section{Conclusions} \label{conclusions}

We applied 3D eclipse mapping techniques to follow the evolution
of the surface brightness of the accretion disc of V4140~Sgr in a
superoutburst. We find that the disc is elliptical in outburst and
decline, with an eccentricity $e=0.13$. In both outburst stages,
the disc orientation is such that superhump maximum occurs when the
secondary star is aligned with the bulge of the elliptical disc.
This lends observational support for the tidal resonance instability
model of superhumps.

The accretion disc fills the primary Roche lobe at outburst 
($R_d=0.85\,R_\mathrm{L1}$), shrinks to $0.6\,R_\mathrm{L1}$ during
decline and reaches a lower value of $0.45\,R_\mathrm{L1}$ in
quiescence. There is marginal evidence that the disc half-opening
angle is larger in outburst ($\beta= 1.0^o\pm 0.5^o$) than in
quiescence ($\beta= 0.5^o\pm 0.5^o$), but the disc is geometrically
thin in all three cases.

The superoutburst occurs at disc temperatures too low to be
accounted for by the disc-instability model even at the upper
$4\,\sigma$ limit on the inferred distance to the binary. The
stability of the eclipse shape, width and depth along 50\,d in
the quiescent period following the superoutburst and the derived 
disc surface brightness distribution (outshining any contribution
from the white dwarf) indicate that the quiescent accretion disc
of V4140~Sgr is in a high-viscosity, steady-state regime.

Flickering mapping of the quiescent data reveal three different
sources of flickering in V4140~Sgr: an azimuthally-extended
stream-disc impact region at disc rim (statistically significant
at the $3\,\sigma$ level) and the innermost disc region
(statistically significant at the $2.5\,\sigma$ level),
responsible for the low-frequency flickering, and an extended,
high-frequency disc component responsible for 2/3 of the total
optical flickering in V4140\,Sgr.
Assuming that the disc-related flickering is caused by fluctuations
in the energy dissipation rate induced by MHD turbulence
according to the model of \cite{ga}, we find that the
quiescent disc viscosity parameter in V4140~Sgr is large
($\alpha \simeq 0.2-0.4$) at all disc radii, in agreement with
MTIM predictions and in marked contrast with DIM predictions.
The discrepancy between the inferred $\alpha$ values and DIM
predictions is statistically significant at the $2\,\sigma$
confidence level outside the region $0.2\leq R/R_\mathrm{L1}\leq 0.3$.

The high-viscosity, steady-state quiescent disc of V4140~Sgr and the
inferred low disc temperatures in superoutburst are inconsistent
with expectations of the disc-instability model, and lead to the
conclusion that the outbursts of V4140~Sgr are powered by
the only other mechanism considered so far, namely,
bursts of enhanced mass transfer rate from its donor star.

The dominant source of uncertainty in the computed $\alpha(R)$
distribution are the errors in the relative amplitude of the
high-frequency, disc-related flickering component. These errors
can be reduced by increasing the $S/N$ of the flickering curves,
i.e., by increasing the statistics and/or the $S/N$ of the
individual light curves included in the data sample.
The benefits of this observational effort are double: it will
allow a more meaningful and statistically robust estimate of
the magnitude of the disc viscosity parameter and it will open
the possibility to properly probe changes in $\alpha$ with radius.

\section{Acknowledgements}

We thank an anonymous referee for useful comments and suggestions
which helped to improve the presentation of our results.
RB acknowledges financial support from CNPq/Brazil through grant 
308\,946/2011-1. ASO acknowledges FAPESP for financial support
under grant 03/12618-7.


\appendix

\section{Performance tests of the 3D eclipse mapping}

Here we report simulations aimed to show the performance of our
3D eclipse mapping code in reproducing accretion disc brightness 
distributions with out-of-eclipse modulations. We also show how
the entropy of the eclipse map can be used to constrain the disc
half-opening angle $\beta$ and the disc rim radius $R_d$.

\begin{figure}
\includegraphics[scale=0.48]{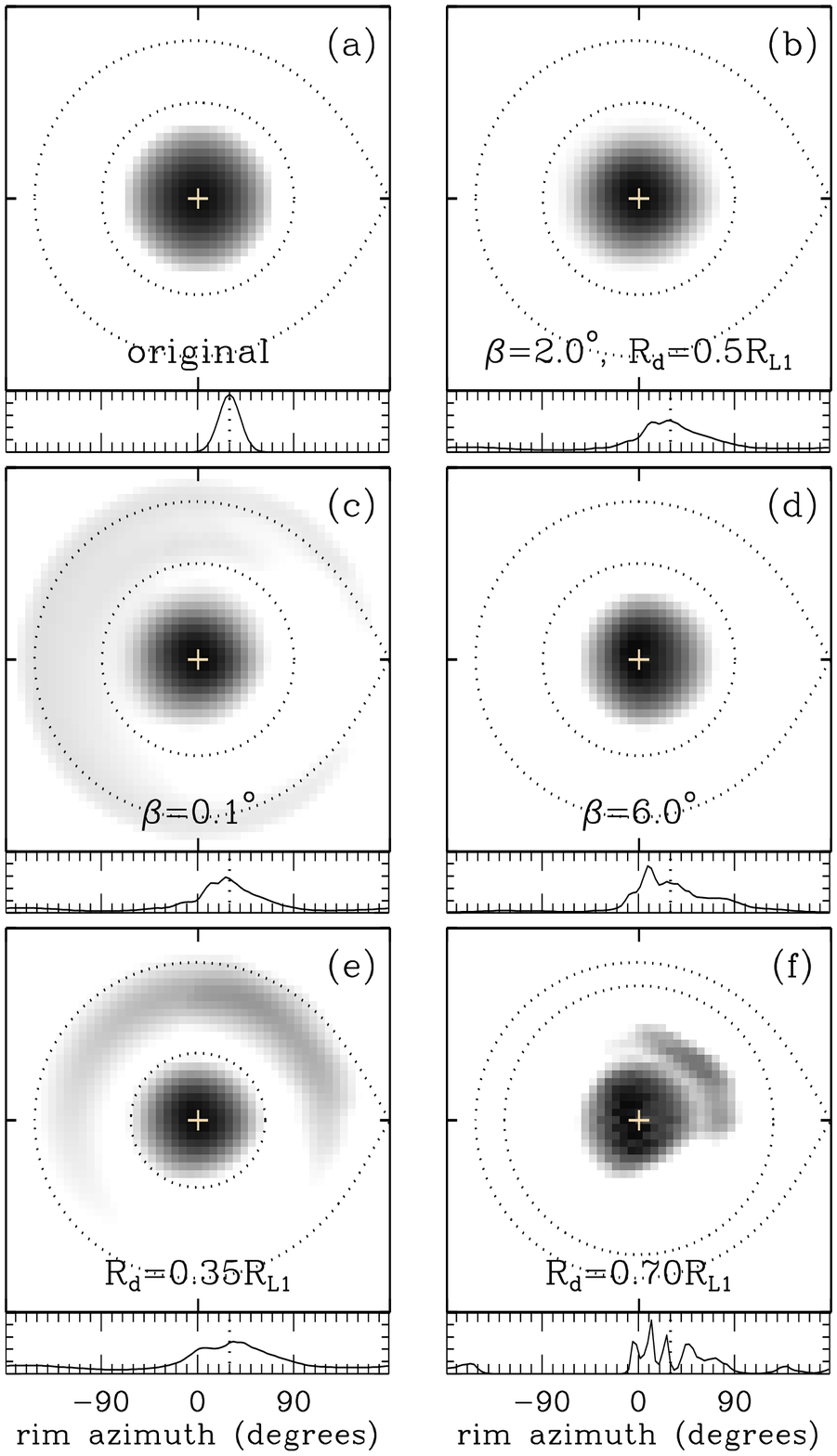}
\caption{ (a): Logarithmic grayscale representation of the synthetic
  brightness distribution used in the simulations (dark regions are
  brighter). A cross marks the disc centre and dotted lines depict
  the primary Roche lobe and the circular disc rim at radius $R_d$.
  The lower inset shows the intensity distribution along the disc rim.
  Azimuths are counted from the line joining both stars and increase
  counter-clockwise. (b) A reconstruction using the correct geometry,
  $\beta=2.0\degr, R_d=0.5\,R_\mathrm{L1}$. (c) A reconstruction with
  correct $R_d$ but underestimated $\beta=0.1\degr$. (d) The same as
  in (c) for an overestimated $\beta=6.0\degr$. (e) A reconstruction
  with correct $\beta$ but underestimated $R_d= 0.35\,R_\mathrm{L1}$.
  (f) The same as in (e) for an overestimated $R_d=0.70\,R_\mathrm{L1}$.
}
\label{figsim3}
\end{figure}

We illustrate the performance of the 3D eclipse mapping code with
reconstructions from a synthetic eclipse map comprised of a
steady-state ($T\propto R^{-3/4}$), axi-symmetric disc brightness 
distribution plus a bright spot ``painted'' in the disc rim at an
azimuth of $\theta_{bs}= +30^o$ (azimuths are counted 
counter-clockwise from the line joining both stars). 
This model distribution is shown in Fig.\,\ref{figsim3}(a).
%
%
It was convolved with the eclipse geometry of V4140~Sgr
($i=80.2^o, q=0.125$) and pairs of values ($\beta,R_d$)
 to generate synthetic light curves to which gaussian noise was added
 in order to simulate real data. Each of these light curves were
 then subjected to the 3D eclipse mapping code.

If an eclipse mapping reconstruction is performed with the wrong choice
of $\beta$ and $R_d$, the code will develop artifacts in the brightness
distribution in order to compensate for the incorrect parameters. This
is illustrated in Fig.\,\ref{figsim3}. A synthetic light curve with
added Gaussian noise ($S/N=50$) was generated from the brightness
distribution of Fig.\,\ref{figsim3}(a) and reconstructions were obtained
for different combinations of $\beta$ and $R_d$. Fig.\,\ref{figsim3}(b)
shows a reconstruction with the correct geometry $\beta=2.0\degr,
R_d=0.5\,R_\mathrm{L1}$. The disc surface brightness distribution and the
azimuth of the spot at disc rim are well recovered. The azimuthal blur
of the spot at disc rim is intrinsic to the eclipse mapping method and
is controlled by the choice of $\Delta\theta$ (see Sect.\,\ref{erup}).
Figs.\,\ref{figsim3}(c) and (d) show the effects of errors in the choice
of $\beta$. Underestimating (overestimating) the disc half-opening
angle forces the code to artificially increase the brightness of the
disc hemisphere farther away from (closest to) the L1 point to
compensate for their lower effective area during eclipse. Eclipse maps
with these spurious, additional structures have lower entropy than the
map with the correct choice of $\beta$ and $R_d$
\footnote{The entropy $S$ measures the amount of structure in an
  eclipse map. The more structured an eclipse map is, the lower
  its entropy is. The map of highest entropy $S_\mathrm{max}$ is the
  one with the least amount of structures allowed by the data. }.
Figs.\,\ref{figsim3}(e) and (f) show the effects of errors in
the choice of $R_d$. Again, artifacts develop in the disc brightness
distribution in order to compensate for the wrong radial position of
the disc rim. In both cases, the additional asymmetry leads to
eclipse maps of lower entropy.
Because of these artifacts, reconstructions obtained with a wrong
choice of ($\beta, R_d$) have lower entropy than that with the correct
combination of parameters.

\begin{figure}
  \includegraphics[bb=0cm 1.5cm 20cm 12cm,scale=0.37, angle=-90]
                  {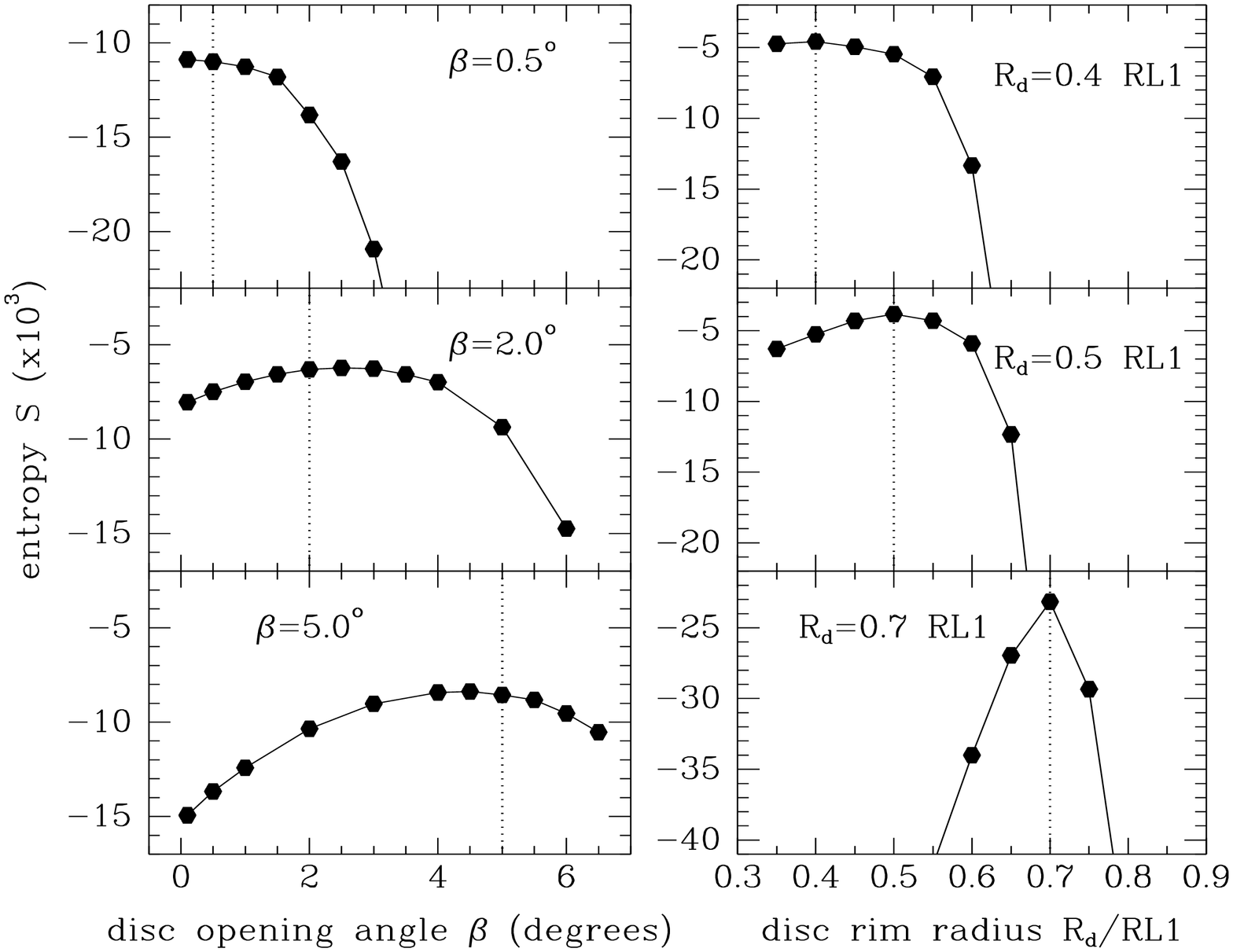}
\caption{ The entropy of the eclipse map as a function of $\beta$
  (left-hand panels) and $R_d$ (right-hand panels). In each panel,
  the correct value of $\beta$ or $R_d$ is labeled and depicted as
  a vertical dotted line.}
\label{figsim2}
\end{figure}

Therefore, it is possible to use the entropy of the eclipse map as a
tool to gauge the correct choice of ($\beta,R_d$). In order to do this,
we obtain reconstructions covering the space of parameters ($\beta,R_d$)
with a suitable sampling and we search for the combination which leads
to the reconstruction of highest entropy ($\beta(S_\mathrm{max}),
R_d(S_\mathrm{max})$). This is called `entropy landscape', a technique
that has been successfully used to find best-fit parameters with other
image reconstruction methods \citep[e.g.,][]{rd94}.
Fig.\,\ref{figsim2} illustrates the procedure.
%
%
We generated an input synthetic light curve ($S/N=50$) from the
brightness distribution of Fig.\,\ref{figsim3}(a) for a given
combination of parameters ($\beta_i,{R_d}_i$) and obtained
reconstructions of that light curve for a range of ($\beta, R_d$)
values. The space of parameters is covered by sampling $\beta$ in
the range $(0\degr - 6\degr)$ at steps of $0.5\degr$, and by
sampling $R_d$ in the range $(0.35-0.85)\,R_\mathrm{L1}$ at steps
of $0.05\,R_\mathrm{L1}$.
The left-hand panels show the entropy of the reconstruction as a
function of $\beta$ for light curves obtained with three different
$\beta_i$ values and fixed $R_d={R_d}_i$, whereas the right-hand
panels show the entropy of the reconstruction as a function of
$R_d$ for light curves obtained with three different ${R_d}_i$
values and fixed $\beta=\beta_i$.
It is possible to recover the correct value of $\beta$ with an
accuracy of $\pm 0.5^o$, and the correct value of $R_d$ with an
accuracy of $\pm 0.05\,R_\mathrm{L1}$. These results are confirmed
with a series of simulations with several different brightness
distributions and light curves with $S/N$ comparable to those of
the V4140~Sgr data.

\bibliographystyle{mnras}

\bsp	
\label{lastpage}
\end{document}